\documentclass[aps,prx,twocolumn,superscriptaddress,export]{revtex4-2} 
\usepackage[colorlinks=true,linkcolor=blue,citecolor=blue,urlcolor=blue]{hyperref}

\usepackage[utf8]{inputenc}
\usepackage[german, english]{babel}

\usepackage{dcolumn}  
\usepackage{bm}        
\usepackage{bbold}
\usepackage{subfigure}
\usepackage{amsthm}
\usepackage{mathtools}
\usepackage{tikz}
\usetikzlibrary{shapes.geometric, arrows, shapes.multipart}
\usepackage{xcolor}
\usepackage{booktabs, tabularx}

\usepackage[export]{adjustbox}
\usepackage{titlesec}
\usepackage{amsfonts}
\usepackage{amssymb}
\usepackage{braket}

\newcommand{\mbf}{\mathbf}

\definecolor{arpit}{RGB}{127,0,0}

\definecolor{dom}{RGB}{34,139,34}

\definecolor{pepe}{rgb}{1,0.49,0}

\definecolor{david}{RGB}{150,0,100}

\definecolor{mike}{RGB}{0,100,100}

\definecolor{ultramarine}{RGB}{0,32,96}

\begin{document}

\title{Fractionalization of subsystem symmetries in two dimensions}
\author{David~T. Stephen}
\affiliation{Department of Physics and Center for Theory of Quantum Matter, University of Colorado, Boulder, CO 80309, USA}
\affiliation{Department of Physics, California Institute of Technology, Pasadena, California 91125, USA}
\author{Arpit Dua}
\affiliation{Department of Physics, California Institute of Technology, Pasadena, California 91125, USA}
\affiliation{Institute for Quantum Information and Matter, California Institute of Technology, Pasadena, California 91125, USA}
\author{Jos\'e Garre-Rubio}
\affiliation{\mbox{University of Vienna, Faculty of Mathematics, Oskar-Morgenstern-Platz 1, 1090 Wien, Austria}}
\author{Dominic~J. Williamson}
\affiliation{Centre for Engineered Quantum Systems, School of Physics, University of Sydney, Sydney, NSW 2006, Australia}
\affiliation{Stanford Institute for Theoretical Physics, Stanford University, Stanford, CA 94305, USA}
\author{Michael Hermele}
\affiliation{Department of Physics and Center for Theory of Quantum Matter, University of Colorado, Boulder, CO 80309, USA}

\date{\today}

\begin{abstract}
The fractionalization of global symmetry charges is a striking hallmark of topological quantum order.
Here, we discuss the fractionalization of subsystem symmetries in two-dimensional topological phases. In line with previous no-go arguments, we show that subsystem symmetry fractionalization is not possible in many cases due to the additional rigid geometric structure of the symmetries. However, we identify a new mechanism that allows fractionalization, involving global relations between macroscopically many symmetry generators. We find that anyons can fractionalize such relations, meaning that the total charge carried under all generators involved in the global relation is non-trivial, despite the fact that these generators multiply to the identity. We first discuss the general algebraic framework needed to characterize this new type of fractionalization, and then explore this framework using a number of exactly solvable models with $\mathbb{Z}_2$ topological order, including models having line and fractal symmetries. These models all showcase another necessary property of subsystem symmetry fractionalization: fractionalized anyons must have restricted mobility when the symmetry is enforced, such that they are confined to a single line or point in the case of line and fractal symmetries, respectively. Looking forward, we expect that our identification of the importance of global relations in fractionalization will hold significance for the classification of phases with subsystem symmetries in all dimensions. 
\end{abstract}

\maketitle

\section{Introduction}

Symmetries that act non-trivially on rigid sub-manifolds of space---subsystem symmetries---are a generalization of global symmetries which have recently come to prominence in a number of diverse settings. For example, generalized gauge theories based on subsystem symmetries lead to exotic fracton topological order \cite{Vijay2016,Williamson2016,Shirley2019},
quantum dynamics with subsystem symmetries can lead to glassy behaviour and anomalous subdiffusive spreading of information \cite{Chamon2005,Haah2013,Prem2017,Iaconis2019}, and field theories having subsystem symmetries display mixing between long range and short range physics (UV/IR mixing) \cite{Seiberg2020,Seiberg2021,Gorantla2021,Distler2022}. On the quantum information side, certain phases of matter protected by subsystem symmetries, such as cluster states, can be used as resources for universal measurement-based quantum computation (MBQC) \cite{Else2012a,Raussendorf2019,Devakul2018a,Stephen2019a,Daniel2019}.
 
In the present work, we are interested in the role that subsystem symmetries play in the classification of topological phases of matter. Subsystem symmetry-protected topological (SSPT) phases of matter---those which are trivial in the absence of symmetry---have by now received considerable attention \cite{Raussendorf2019,You2018,Devakul2018,Devakul2019,Devakul2019a,Tantivasadakarn2019,You2020}, due in part to their aforementioned ability to perform universal MBQC and their unique entanglement properties \cite{Williamson2019,Stephen2019,Schmitz2019,SanMiguel2021}. On the other hand, subsystem symmetry-enriched topological (SSET) phases, where the symmetry fractionalizes on non-trivial bulk excitations of a topologically ordered system, remain largely unexplored.

In Ref.~\cite{Stephen2020}, it was argued that subsystem symmetry fractionalization cannot occur in fewer than three spatial dimensions (3D). The basic argument is as follows. Fractionalization of global symmetries occurs when a point-like topological excitation (anyon) carries a fractional charge under a certain symmetry operator. Fractional charges are possible because there are global conservation laws on the number of anyons in the system which ensure that the total charge of all excitations always forms a linear representation of the global symmetry. For example, in the $\nu=\frac{1}{3}$ fractional quantum hall effect, anyons come in multiples of three, so they are allowed to carry $\frac{1}{3}$ of the elementary charge \cite{Laughlin1983}.
For subsystem symmetries, however, there is no constraint on the number of anyons acted on by a given symmetry generator. This is because the anyons of a general topological order in two spatial dimensions (2D) can be moved freely throughout the system, and in particular can be freely moved in or out of a given subsystem. For this reason, it appears as though a single anyon always transforms under a linear representation of a subsystem symmetry, and therefore cannot carry fractional subsystem symmetry charge in 2D.

In 3D and higher, there are several ways around this argument. For example, fracton topological order has topological excitations with restricted mobility that can be described in terms of conservation laws on rigid subsystems, allowing fractionalization of symmetries that act on the same subsystems \cite{You2020}. Also, topological orders in 3D and higher can have higher dimensional excitations, such as the loop-like excitations in the 3D toric code, which always \textit{e.g.} penetrate any plane an even number of times, giving a sort of generalized conservation law that again allows fractionalization of planar subsystem symmetries on the loop excitations \cite{Stephen2020}.

In this paper, we show that subsystem symmetry fractionalization is, in fact, possible in 2D. We take advantage of two key observations that the previous argument does not account for. First, it is possible that the mobility of anyons becomes restricted when subsystem symmetries are enforced, giving rise to emergent conservation laws on subsystems in the presence of symmetry. This violates the assumption in the previous argument that anyons can be freely moved in or out of a given subsystem. Second, we find that this mobility restriction enables a novel type of fractionalization that is unique to subsystem symmetries, which involves global relations between macroscopically many symmetry generators. As a concrete example, if the subsystem symmetry is generated by operators that flip all spins along a given row or column of a square lattice, then these symmetries satisfy a global relation which says that the product of generators acting on all rows and columns is the identity operator. Despite this, we find that anyons can carry a non-zero total charge under the product of all generators, meaning that this global relation is fractionalized. We furthermore argue that this is the only type of subsystem symmetry fractionalization that is possible in 2D by formalizing the argument given above.

\begin{figure}[t]
\centering
\subfigure[]{{\label{fig:localization}\includegraphics[width=\linewidth]{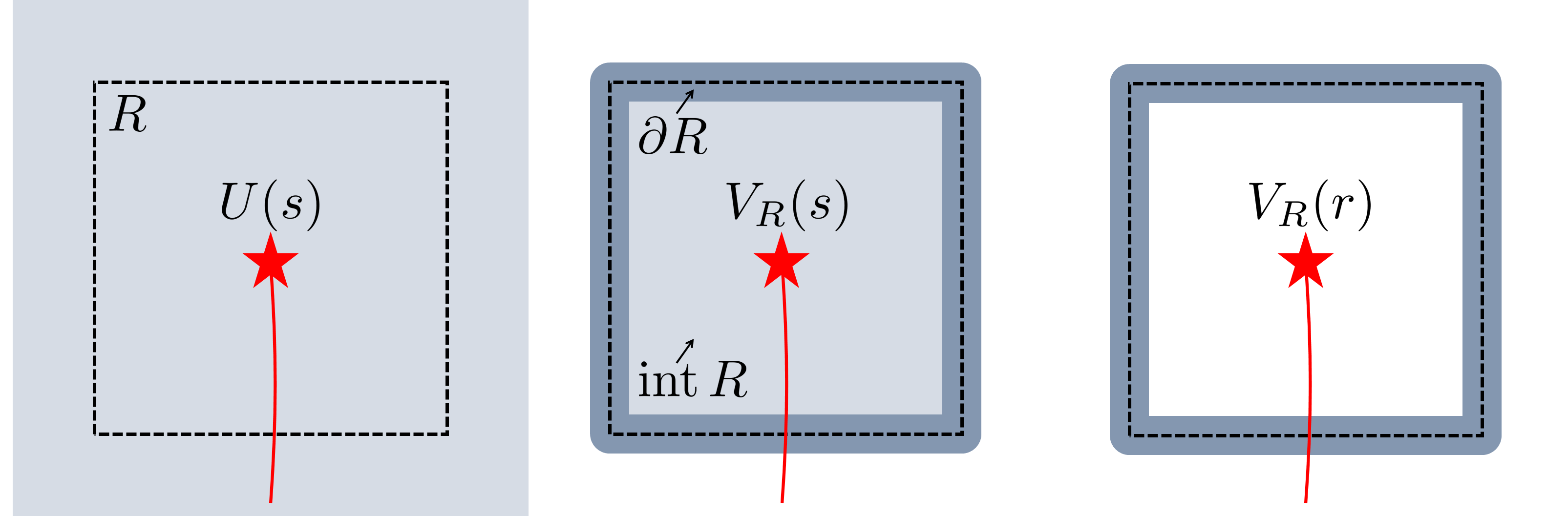}}} \hfill
\subfigure[]{\label{fig:relations}\includegraphics[width=\linewidth]{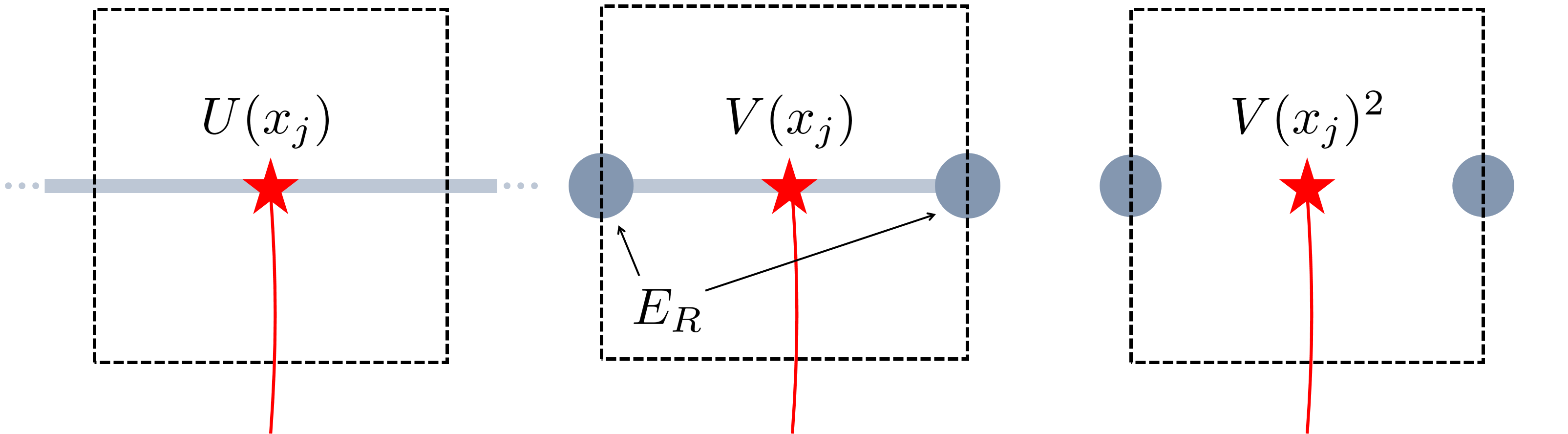}}
\subfigure[]{\label{fig:relations_all}\includegraphics[width=\linewidth]{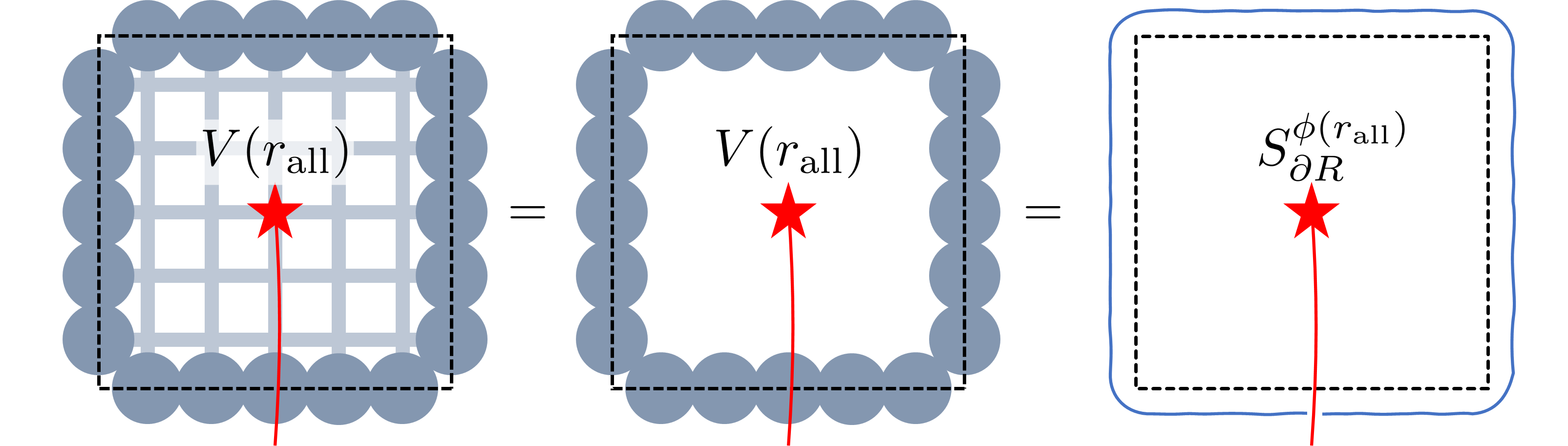}}
\caption{(a) Localization of a global symmetry. Left: Action of a global symmetry $U(s)$ on a localizable state containing an anyon (star) inside a finite rectangular region $R$ indicated by dotted lines. Middle: The symmetry can be localized to $R$ using an operator $V_R(s)$ that acts only in the vicinity of $R$. In the interior of $R$ ($\operatorname{int} R$, light shaded region), $V_R(s)$ acts as $U(s)$. Near the boundary of $R$ ($\partial R$, dark shaded region), $V_R(s)$ in general acts different from $U(s)$. Right: A relation $r$ localized to $R$ gives an operator $V_R(r)$ which necessarily acts trivially in $\operatorname{int} R$, but may act non-trivially on $\partial R$ and intersect the string operators of anyons contained within $R$. (b) Localization of a subsystem symmetry. Left: A subsystem symmetry $U(x_j)$ supported on a horizontal line that crosses $R$. Middle: The symmetry can be localized by acting only near the area where the symmetry crosses $\partial R$. Right: The operator $V(x_j)^2$ acts non-trivially only near the left and right boundaries of $R$ such that it does not touch a string operator that exits $R$ through the bottom. (c) Localization of a fractionalizable relation. Left: The product of all localized row and column symmetries supported within $R$ gives $V(r_{\mathrm{all}})$. Middle: The bulk symmetry action cancels out, leaving only a boundary action. Right: This boundary action is equivalent to a string operator $S^{\phi(r_{\mathrm{all}})}_{\partial R}$ which braids some abelian anyon $\phi(r_{\mathrm{all}})$ around $R$, up to corner effects.}
\label{fig:fractionalization}
\end{figure}

To characterize subsystem symmetry fractionalization in general, we work with symmetry-localized operators, which are restrictions of the symmetry to a finite region, such that no excitations are created on the region's boundary. Given a relation in the symmetry group, the corresponding product of symmetry-localized operators multiplies to the identity in the bulk of the region. However, such a product may act non-trivially on the boundary. In particular, for a global relation between subsystem symmetries, this boundary action is equivalent to the action of a string operator that creates a pair of anyons, braids them around the boundary of the region, and annihilates them back to vacuum, see Fig.~\ref{fig:fractionalization}. In this way, we uncover a framework for labelling different classes of subsystem symmetry fractionalization in terms of maps between a group of global relations among symmetry generators and the fusion group of abelian anyons of the topological order.


After introducing this general framework (Section \ref{sec:fractionalization}), we explore it with a number of exactly solvable lattice models having $\mathbb{Z}_2$ topological order. We first consider the example of row and column symmetries described above, where we give a model realizing each fractionalization class that is possible in our general framework (Section \ref{sec:line}). In this case, a fractionalization class is labeled by a single anyon type. We show that all particles that braid non-trivially with the labeling anyon become \textit{symmetry-protected lineons}, which can only be moved along either horizontal or vertical lines under dynamics respecting the symmetry.
We then consider a larger subsystem symmetry group which adds symmetries acting along diagonal lines (Section \ref{sec:threelines}). This results in a larger group of global relations, increasing the number of possible fractionalization classes. As a final example, we consider fractal subsystem symmetries (Section \ref{sec:fractal}). We show that the same concepts of global relations characterize fractionalization here as well, and that fractionalized anyons become \textit{symmetry-protected fractons} which appear at the corners of fractal operators and are immobile when symmetry is enforced. Finally, we briefly discuss more general settings where subsystem symmetry fractionalization can occur, and highlight further directions of future research (Section \ref{sec:outlook}).

\section{Fractionalization of subsystem symmetries} \label{sec:fractionalization}

In this section we develop a framework to describe subsystem symmetry fractionalization, which allows us to understand what kinds of fractionalization are possible. We focus throughout on two-dimensional gapped systems, and in particular on topologically ordered systems, \emph{i.e.} those with anyon excitations. We show that a more precise version of the argument given in the introduction does indeed constrain what types of fractionalization are possible, but also that opportunities for non-trivial fractionalization still remain.

\subsection{The case of global symmetries}

We first give an account of symmetry fractionalization for global symmetries \cite{Laughlin1983,Wen2002,Essin2013,Mesaros2013,Chen2015,Barkeshli2019}, suitable for generalization to subsystem symmetries. Throughout the paper, by global symmetries we mean more specifically unitary internal zero-form symmetries. While the results reviewed are well-known, some features of our treatment -- in particular, the association between localized relations in the symmetry group and anyon string operators -- have not appeared previously in the literature to our knowledge.

Let $G$ be the symmetry group, and let $U(s)$ be a representation acting on the Hilbert space, such that $U(g_1) U(g_2) = U(g_1 g_2)$ for any group elements $g_1$ and $g_2$. A key ingredient for symmetry fractionalization is the concept of symmetry localization. Consider a region of space $R$ with the topology of a disc, and with characteristic linear size $\ell_R$ much larger than the correlation length of the system. We also define the boundary and interior of $R$ as $\partial R$ and $\operatorname{int} R = R - R \cap \partial R$ as shown in Fig.~\ref{fig:fractionalization}. The thickness $\ell_{\partial R}$ of the annulus $\partial R$ satisfies $\ell_{\partial R} \ll \ell_R$, while still being much greater than the correlation length. We are interested in localizable states that ``look locally like the ground state'' within $\partial R$, but may contain excitations in $\operatorname{int} R$ or outside $R \cup \partial R$. More precisely, localizable states are those whose reduced density matrices on all disc-shaped regions within $\partial R$ are identical to those of a ground state.

We now define an operator $V_R(g)$ supported on $R \cup \partial R$ that localizes the symmetry $U(g)$ to the region $R$. Intuitively, $V_R(g)$ acts like $U(g)$ within $\operatorname{int} R$, while creating no excitations in $\partial R$. To construct $V_R(g)$, we observe that $U(g)$ is a product of on-site unitaries, and by taking the product only over sites within $R$, we obtain a truncated symmetry operator $U_R(g)$ supported on $R$. This is not generally a localization of the symmetry, because acting with $U_R(g)$ on a localizable state typically creates excitations within $\partial R$.

To construct $V_R(g)$, we multiply $U_R(g)$ by a local unitary (finite-depth circuit) supported on $\partial R$, whose effect is to ``clean up'' the excitations created by $U_R(g)$. This can only be done if the domain wall excitation created at the boundary of $R$ is in a trivial superselection sector. For instance, we cannot construct $V_R(g)$ if the symmetry is spontaneously broken. An obstruction also arises when anyon excitations are present, if the symmetry $g$ realizes a non-trivial permutation of anyon types; if $R$ contains an anyon whose type is changed by $g$, then acting with $U_R(g)$ leaves behind an anyon in $\partial R$, which cannot be removed by acting with a local unitary supported on $\partial R$. However, it is believed that the symmetry localization $V_R(g)$ always exists for unbroken symmetries that do not permute anyon types.

Multiplying two symmetry localizations results in
\begin{equation}
V_R(g_1) V_R(g_2) = \omega(g_1, g_2) V_R(g_1 g_2) \text{,}
\end{equation}
where $\omega(g_1, g_2)$ is an operator supported on $\partial R$. This and other similar equations hold on the subspace of all localizable states, and on this subspace the only relevant data about $\omega(g_1, g_2)$ is its type as an anyon string operator. In fact, $\omega(g_1, g_2)$ must be the string operator for an abelian anyon \cite{Chen2015,Barkeshli2019}. We note that, on the subspace of localizable states, string operators for abelian anyons must commute with the symmetry localizations $V_R(g)$, because otherwise $V_R(g)$ can change the anyon type contained in $R$. From this fact and associativity of multiplication of the $V_R(g)$'s, we obtain
\begin{equation}
\omega(g_1, g_2) \omega(g_1 g_2, g_3) = \omega(g_1, g_2 g_3) \omega(g_2, g_3) \text{,}
\end{equation}
which is the so-called 2-cocycle condition on $\omega$. We therefore think of $\omega$ as a 2-cocycle valued in the fusion group of abelian anyons ${\cal A}$.

Finally, we account for an ambiguity in the construction of $V_R(g)$. Namely, we are free to modify $V_R(g) \to \lambda(g) V_R(g)$, where $\lambda(g)$ is an (abelian) anyon string operator supported on $\partial R$. Such modifications are allowed because $V_R(g)$ and $\lambda(g) V_R(g)$ are equally good localizations of the symmetry $g$. Under such a transformation of $V_R(g)$, the 2-cocycle $\omega$ is multiplied by a 2-coboundary,
\begin{equation}
    \omega(g_1,g_2)\to \frac{\lambda(g_1)\lambda(g_2)}{\lambda(g_1g_2)}\omega(g_1,g_2).
\end{equation}
However the cohomology class $[\omega] \in H^2(G, \mathcal{A})$, defined as the group of cocycles modulo such coboundary transformations, is unchanged. This cohomology class labels distinct types of symmetry fractionalization for a given topological order, and is often referred to as a fractionalization class in this context.

When we discuss subsystem symmetries, it is convenient to work with generators and relations for the symmetry group. To that end, we first discuss how fractionalization of a global symmetry can be described in terms of generators and relations. A general symmetry group $G$ can be described by a set of generators $s\in \mathcal{S}$ and a set of relations $r\in \mathcal{R}$ among them, \textit{i.e.} products of the generators (and their inverses) that equal the identity element. More formally, we express $G = F(\mathcal{S})/\mathfrak{R}$, where $F(\mathcal{S})$ is the free group over the set of generators, and $\mathfrak{R}$ is a \emph{group of relations} defined as the smallest normal subgroup of $F(\mathcal{S})$ containing $\mathcal{R}$. That is, the group $\mathfrak{R}$ can be generated by taking arbitrary products of elements of  $\mathcal{R}$ and their inverses, and also by conjugating elements of $\mathcal{R}$ by arbitrary generators. Given a representation of the group on Hilbert space $U(g)$, for any relation $s_1 s_2 \cdots s_n = 1$ (or more generally any element of $\mathfrak{R})$, we have $U(s_1) U(s_2) \cdots U(s_n) = 1$.



For some relation $r = s_1 \cdots s_n = 1$, the operator $V_R(r) \equiv V_R(s_1) \cdots V_R(s_n)$ is supported only on $\partial R$, and may be a string operator of a non-trivial abelian anyon. This is potentially a sign of symmetry fractionalization; if the anyon string type on $\partial R$ is $a$, and an anyon of type $b$ lies inside $\operatorname{int} R$, then the relation $r$ acts on the $b$-anyon as the phase factor $e^{i \Theta_{a b}}$, where $\Theta_{a b}$ is the mutual statistics angle between $a$ and $b$. However, care must be taken in concluding non-trivial symmetry fractionalization is present, because in general the anyon string type of $V_R(r)$ can change if the $V_R(s_i)$ are redefined according to the coboundary transformations discussed above.

More generally, we have a homomorphism $\phi : \mathfrak{R} \to {\cal A}$. For $r \in \mathfrak{R}$, $\phi(r)$ is the anyon type of the string appearing in $\partial R$ for the operator $V_R(r)$. The values $\phi(r)$ in general depend on redefinitions of the $V_R(s)$ as noted above. In Appendix~\ref{app:omega}, we explain how the homomorphism $\phi$ gives a 2-cocycle, and thus specifies a fractionalization class $[\omega] \in H^2(G, {\cal A})$.

\subsection{The case of subsystem symmetries}

Now we investigate how the above discussion changes when we consider subsystem symmetries. As a concrete example to guide our discussion, we consider the case of line-like subsystem symmetries on an infinite 2D square lattice. We first write down the subsystem symmetry group as an abstract group without specifying the geometry of the generators. The group $G$ is generated by the elements $x_j$ and $y_i$ where $i,j\in \mathbb{Z}$ (which correspond to symmetry actions on row $j$ and column $i$) subject to the following relations,
\begin{equation} \label{eq:relations}
\begin{aligned}
    x_j^2 &= 1 \qquad \forall j\in \mathbb{Z} ,\\
    y_i^2 &= 1 \qquad \forall i\in \mathbb{Z} ,\\
    x_ix_jx_i^{-1}x_j^{-1} &=  1 \qquad \forall i,j\in \mathbb{Z} ,\\
    y_iy_jy_i^{-1}y_j^{-1} &= 1 \qquad \forall i,j\in \mathbb{Z} ,\\
    x_jy_ix_j^{-1}y_i^{-1}& = 1 \qquad \forall i,j\in \mathbb{Z} ,\\
    \prod_j x_j \prod_i y_j^{-1} &= 1 
    \, .
\end{aligned}
\end{equation}
Here, the final relation corresponds to the fact that the product of all lines in both directions is the identity. It is a global relation, we which define as a relation involving macroscopically many generators, \textit{i.e.} the number of involved generators grows (sub)extensively with the system size. In principle, it is possible that any of the above relations can fractionalize when acting on an isolated anyon. However, we have found that the geometric constraints of line-like symmetries make fractionalization impossible for all but the final relation.

To introduce geometry explicitly, suppose the generators $U(x_j)$ and $U(y_i)$ act only on the sites within row $j$ and column $i$, respectively. Let the region $R$ be a finite rectangle that is much larger than the correlation length of the system. We assume the symmetry localization of $U(s)$ to $R$ exists for each generator $s$, and we denote the symmetry localization by $V(s)$ instead of $V_R(s)$ to simplify the notation. At the end of this section, we discuss the physical meaning of assuming that symmetry localizations exist.

As in the case of global symmetry, we construct $V(s)$ by first truncating $U(s)$ to $U_R(s)$ with support in $R$. When $U_R(s)$ acts on a localizable state, it can only create excitations within a region $E_R(s)$, which consists of the union of all small discs in $\partial R$ in which $U_R(s)$ differs from $U(s)$. We note that $E_R(s) \subset \operatorname{Supp} U(s) \cap \partial R$, where $\operatorname{Supp} U(s)$ is the support of $U(s)$, thickened by an amount on the scale of the correlation length. For subsystem symmetries, in contrast to global symmetries, $E_R(s)$ is generally not all of $\partial R$.
We obtain $V(s)$ from $U_R(s)$ by multiplying with a local unitary supported on $E_R(s)$. This is illustrated in Fig.~\ref{fig:relations} for $V(x_j)$, where $E_R(x_j)$ consists of two disjoint discs where row $j$ intersects $\partial R$. We see right away that $V(x_j)$ cannot be multiplied by an anyon string operator while maintaining its support, because such a string is supported on all of $\partial R$. It thus appears that there is no important ambiguity---and certainly no ambiguity involving the underlying topological order---involved in constructing symmetry localizations for the generators of a linear subsystem symmetry. As a consequence of this lack of ambiguity we see below that distinct types of subsystem symmetry fractionalization are not labeled by cohomology classes.

Now we consider the relation $x_j^2 = 1$. The product $V(x_j)^2$ has support only in two disjoint localized discs on the left and right edges of $R$. Similar to the above discussion on the lack of ambiguity in constructing $V(x_j)$, there is thus no way to associate a string operator on $\partial R$ with the relation $x_j^2 = 1$. Moreover, if $R$ contains an anyon, we can always direct the associated string operator out of $R$ through the top or bottom boundaries, such that this string is acted on trivially by $V(x_j)^2$, as shown in Fig.~\ref{fig:relations}. Therefore, this relation cannot be fractionalized. The same argument rules out fractionalization of the third relation in Eq.~(\ref{eq:relations}), and for the second and fourth if we swap left and right with top and bottom. For the fifth relation, notice that $V(x_j)$ and $V(y_i)$ commute with each other since $E_R(x_j)$ and $E_R(y_j)$ do not intersect, except possibly near a corner of $R$, if row $j$ and column $i$ lie near the edges of $R$. In that case, we can route an anyon string out of $R$ while avoiding the corners, implying there is no fractionalization again.

For the final relation, which we denote as $r_{\mathrm{all}}$ since it is a product of all generators, we cannot use the above geometric arguments because the symmetry operators involved are in fact global symmetries. That is, $\prod_j U(x_j)$ and $\prod_i U(y_i)$ act non-trivially on the entire region $R$, so in general $V(r_{\mathrm{all}})  \equiv \prod_j V(x_j) \prod_i V(y_i)^{-1}$ acts on the entire boundary of $R$. If we multiply all truncated line symmetries $V(x_j)$ and $V(y_i)$ within $R$, the symmetry action in the bulk necessarily multiplies to the identity, but the boundary actions from each line may multiply to a non-trivial operator, as shown in Fig.~\ref{fig:relations_all}. We find that, interestingly, such a boundary operator can be equivalent to braiding some anyon around $R$, as in the case of conventional symmetry fractionalization. More generally, we see that only some relations $r \in \mathcal{R}$ are \emph{fractionalizable}, so we partition $\mathcal{R}$ into disjoint subsets of fractionalizable relations ($\mathcal{R}_f$) and non-fractionalizable relations ($\mathcal{R}_{nf}$). We say a relation $r=s_1\dots s_n$ is fractionalizable if $E_R(r) \equiv \bigcup_i E_R(s_i) \supset \partial R$, so that $V(r)$ has the support necessary to contain a string operator. A fractionalizable relation must necessarily be a global relation, but in general not every global relation is fractionalizable \footnote{For example, if we consider a group generated by pairs of row symmetries $x_jx_{j+1}$, then the product of all pairs is a global relation $r$ that is not fractionalizable since $E_R(r)$ consists of only the left and right edges of $R$.}.

We therefore propose that symmetry fractionalization for a general subsystem symmetry is characterized by a homomorphism $\phi : \mathfrak{R}_f \to \mathcal{A}$, where $\mathfrak{R}_f$ is an abelian group of fractionalizable relations that we define precisely below. Roughly speaking, $\mathfrak{R}_f$ is the set of fractionalizable relations $\mathcal{R}_f$, made into an abelian group. For $r \in \mathcal{R}_f$, the homomorphism $\phi(r)\in \mathcal{A}$ is the anyon whose string realizes the boundary action of the relation $r$ on a finite region, as in the fractionalization of global symmetries described above. We often refer to the value of $\phi(r)$ as the fractionalization of the relation $r$. One can also view $\phi(r)$ as a generalization of a twist defect for relations, which appears at the endpoints of the boundary action $V(r)$ when it is cut open \cite{Bombin2010,tarantino2015symmetry,Barkeshli2019}.
For the line symmetries described above, we have only a single fractionalizable relation $r_{\mathrm{all}}$, and $\mathfrak{R}_f \cong \mathbb{Z}_2$, but more complicated groups of fractionalizable relations are possible for different geometries of subsystem symmetry, as we demonstrate in the examples below. In this language, the information about the topological order enters as the group $\mathcal{A}$, while the algebraic and geometric information of the subsystem symmetries enters through $\mathfrak{R}_f$.

To define the group of fractionalizable relations more formally, we let $\mathfrak{R}_{nf}$ be the smallest normal subgroup of $F(\mathcal{S})$ containing the set of non-fractionalizable relations $\mathcal{R}_{nf}$.  Then we define $\mathfrak{R}_f = \mathfrak{R} / \mathfrak{R}_{nf}$. In all the examples considered in this paper, $\mathfrak{R}_f$ thus defined is always abelian. Moreover, a homomorphism $\phi : \mathfrak{R}_f \to {\cal A}$ can always be extended to a homomorphism $\phi : \mathfrak{R} \to {\cal A}$ which is trivial on $\mathfrak{R}_{nf}$. Therefore a homomorphism $\phi : \mathfrak{R}_f \to {\cal A}$ defines a 2-cocycle on the group of subsystem symmetries valued in ${\cal A}$.

Now we are in a position to take note of two important differences from the fractionaliation of a global symmetry. First, because $\phi$ is trivial on $\mathfrak{R}_{nf}$, only a certain subgroup of 2-cocycles can be realized. Second, there are no coboundary transformations, due to the lack of ambiguity in constructing symmetry localizations discussed above. Therefore we conjecture that different types of subsystem symmetry fractionalization are classified by the subgroup of 2-cocycles that can be realized by homomorphisms $\phi: \mathfrak{R}_f \to {\cal A}$. In fact, this statement is a slight oversimplication; if two homomorphisms $\phi$ (or equivalently their corresponding 2-cocycles) are related by a relabeling of anyons that is a symmetry of the underlying unitary modular tensor category -- such as relabeling $e \leftrightarrow m$ in the toric code -- then the corresponding types of fractionalization should be considered equivalent.

We conclude this section by discussing the physical meaning of assuming that the symmetry localization $V(s)$ exists for all generators of the subsystem symmetry. It may appear we are thus assuming that the symmetry does not permute anyons, but we argue in Appendix~\ref{app:permute} that subsystem symmetries in fact cannot permute anyons. However, we are still making a non-trivial assumption, because there are models where symmetry localizations do not exist. A simple example is obtained by taking the usual 2D toric code on the square lattice \cite{Kitaev2003}, and considering a linear subsystem symmetry generated by rigid string operators running along the $x$- and $y$-directions; for concreteness, we consider string operators that transport vertex excitations, which we refer to as $\mathrm{e}$-particles. If such a string operator $S$ cuts through a rectangular region $R$, the truncated operator $S_R$ creates two $\mathrm{e}$ particles at the points where $S$ intersects $\partial R$, and these non-trivial excitations cannot be removed by modifying $S_R$ near its two endpoints.

In the above toric code example, we view the subsystem symmetry as being spontaneously broken. First we recall that the toric code has a spontaneously broken one-form symmetry, of which the subsystem symmetry discussed above is a ``subgroup.'' Moreover, on a space with periodic boundary conditions, the subsystem symmetry acts non-trivially within the degenerate space of ground states, which is a diagnostic of symmetry breaking that applies both to conventional global symmetries and higher-form symmetries.

In fact, we propose that spontaneous symmetry breaking for subsystem symmetries can be \emph{defined} in terms of symmetry localization, at least in gapped systems. Namely, we say that a subsystem symmetry is preserved if symmetry localizations exist for all its generators. Otherwise, we say the symmetry is broken. This definition is motivated in part by the fact that breaking of a global symmetry is an obstruction to symmetry localization, because the restriction of a broken global symmetry to a region creates a non-trivial domain wall on the boundary. This issue plays an important role in our treatment of a system with fractal subsystem symmetry, and it is discussed further in that context (Section~\ref{sec:fractal}).

\section{Examples with line symmetries} \label{sec:line}

In this section, we present explicit examples of Hamiltonians with 2D $\mathbb{Z}_2$ topological order which respect line symmetries along two orthogonal directions, as described above. After recalling the basic properties of the anyons in $\mathbb{Z}_2$ topological order, we introduce examples where $\phi(r_{\mathrm{all}}) = a$ for each anyon $a$.

\subsection{$\mathbb{Z}_2$ topological order}

Most of the models discussed in the paper support $\mathbb{Z}_2$ topological order, so we briefly recall its basic properties \cite{Kitaev2003}. The theory has four distinct topological sectors labelled by $1$, $\mathrm{e}$, $\mathrm{m}$, and $\epsilon=\mathrm{e}\times\mathrm{m}$. $1$-particles correspond to the topologically trivial excitations. The three non-trivial excitations, called anyons, each square to $1$, each have a mutual $-1$ braiding statistic with the other two anyons, and have fusion rules such that fusing any two distinct anyons gives the third. These fusion rules are encoded in the group ${\cal A} = \mathbb{Z}_2 \times \mathbb{Z}_2$. The anyons $\mathrm{e}$ and $\mathrm{m}$ exhibit bosonic self-statistics while $\epsilon$ exhibits fermion self-statistics.

A simple model that captures $\mathbb{Z}_2$ topological order is the toric code Hamiltonian \cite{Kitaev2003}. To write down explicit commuting projector Hamiltonians, we work with models on a square lattice with qubits assigned to the sites and edges. The standard toric code on the square lattice has only edge qubits; we also include the site qubits to match the degrees of freedom of the upcoming models with fractionalized subsystem symmetry. We label sites by an index $v=(i,j)$ and define the unit vectors $\hat{x}=(1,0)$ and $\hat{y}=(0,1)$. We also make use of half coordinates to label edges, such that the edge north/south (east/west) of site $v$ is labelled $v\pm \frac{\hat{y}}{2}$ ($v\pm\frac{\hat{x}}{2}$). The toric code Hamiltonian is defined as,
\begin{equation}
    H_{TC} = - \sum_v X_v - \sum_v A_v - \sum_p B_p
\end{equation}
where,
\begin{equation}
    A_v = \prod_{e\ni v} Z_e,
\end{equation}
and,
\begin{equation}
    B_p = \prod_{e\in p} X_e,
\end{equation}
where $e\ni v$ denotes the four edges incident on site $v$, $e\in p$ denotes the four edges around a plaquette $p$, and $X$ and $Z$ are the standard qubit Pauli operators. Finally, the term $-\sum_v X_v$ puts the site qubits into the trivial product state $\bigotimes_v |+\rangle_v $, where $|+\rangle = \frac{1}{\sqrt{2}}(|0\rangle + |1\rangle)$.

In $H_{TC}$, the $\mathrm{e}$-anyons and $\mathrm{m}$-anyons correspond to violations of $A_v$ and $B_p$, respectively, which are created at the endpoints of string operators,
\begin{equation}
    S^{\mathrm{e}}_\gamma = \prod_{e\in\gamma} X_e, \qquad
    S^{\mathrm{m}}_\lambda = \prod_{e\in\lambda} Z_e
\end{equation}
where $\gamma$ ($\lambda$) is a connected line of edges on the (dual) square lattice. The fermion $\epsilon$ corresponds to a violation of both $A_v$ and a nearby $B_p$. The braiding and fusion rules of the anyons, as described above, can be verified by the algebraic properties of these string operators. 

We consider the linear $\mathbb{Z}_2$ subsystem symmetry described above in Sec.~\ref{sec:fractionalization}, with generators $x_j$ and $y_i$ corresponding to the action of symmetry on rows and columns, respectively. The symmetry acts on the quantum degrees of freedom by the representation,
\begin{equation} \label{eq:line_symms}
    U(x_j) = \prod_{i} X_{(i,j)}, \quad U(y_i) = \prod_{j} X_{(i,j)}.
\end{equation}
$H_{TC}$ trivially commutes with these symmetries. The symmetries do not act on the anyons in $H_{TC}$, so there is no subsystem symmetry fractionalization in this model. This is clear because the symmetry only acts on the site qubits, which are decoupled from the edge qubits forming the $\mathbb{Z}_2$ topolocally ordered state.

\subsection{Model with subsystem symmetry fractionalization} \label{sec:hl}

\begin{figure}[t]
\centering
\subfigure[]{{\label{fig:hl_terms}\includegraphics[scale=0.4]{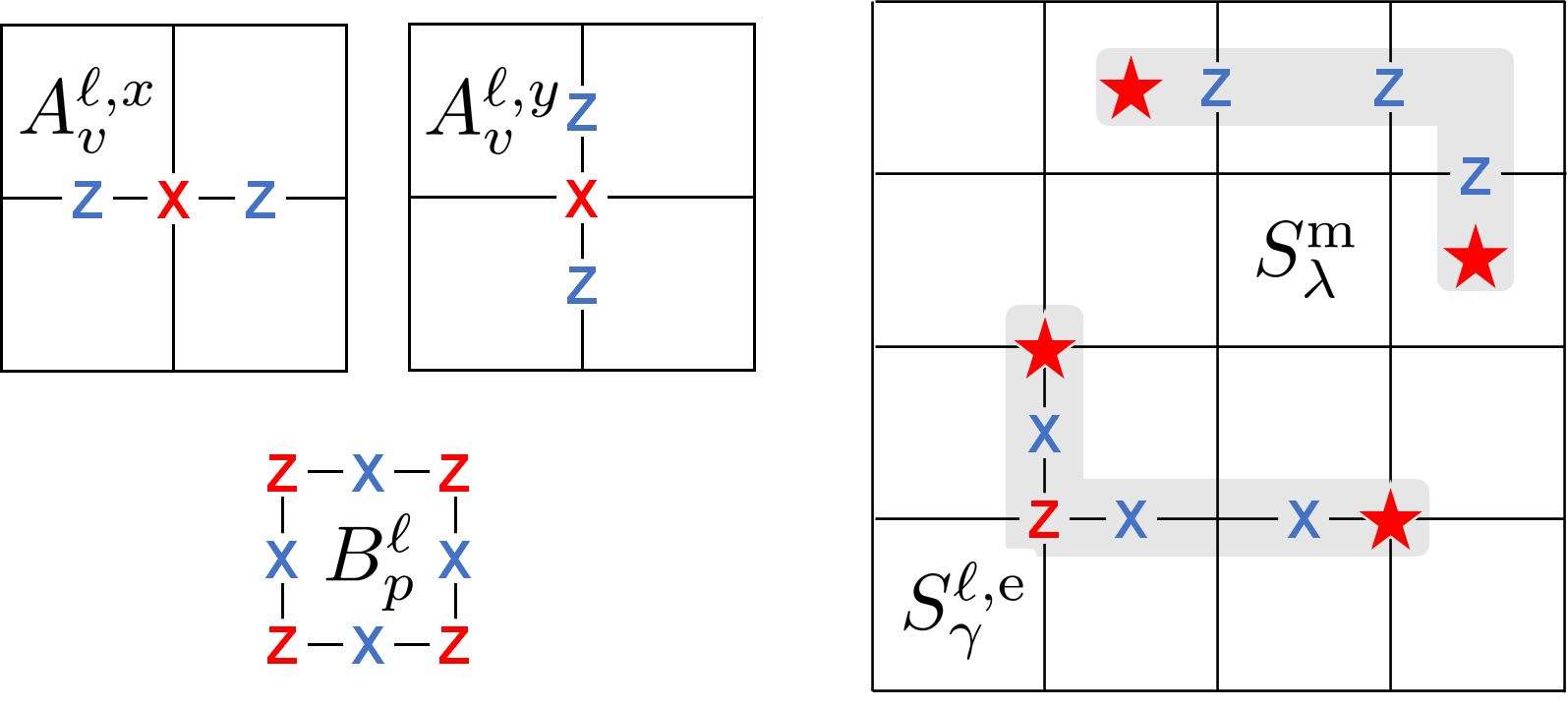}}} \hfill
\subfigure[]{{\label{fig:hl_rel}\includegraphics[scale=0.4]{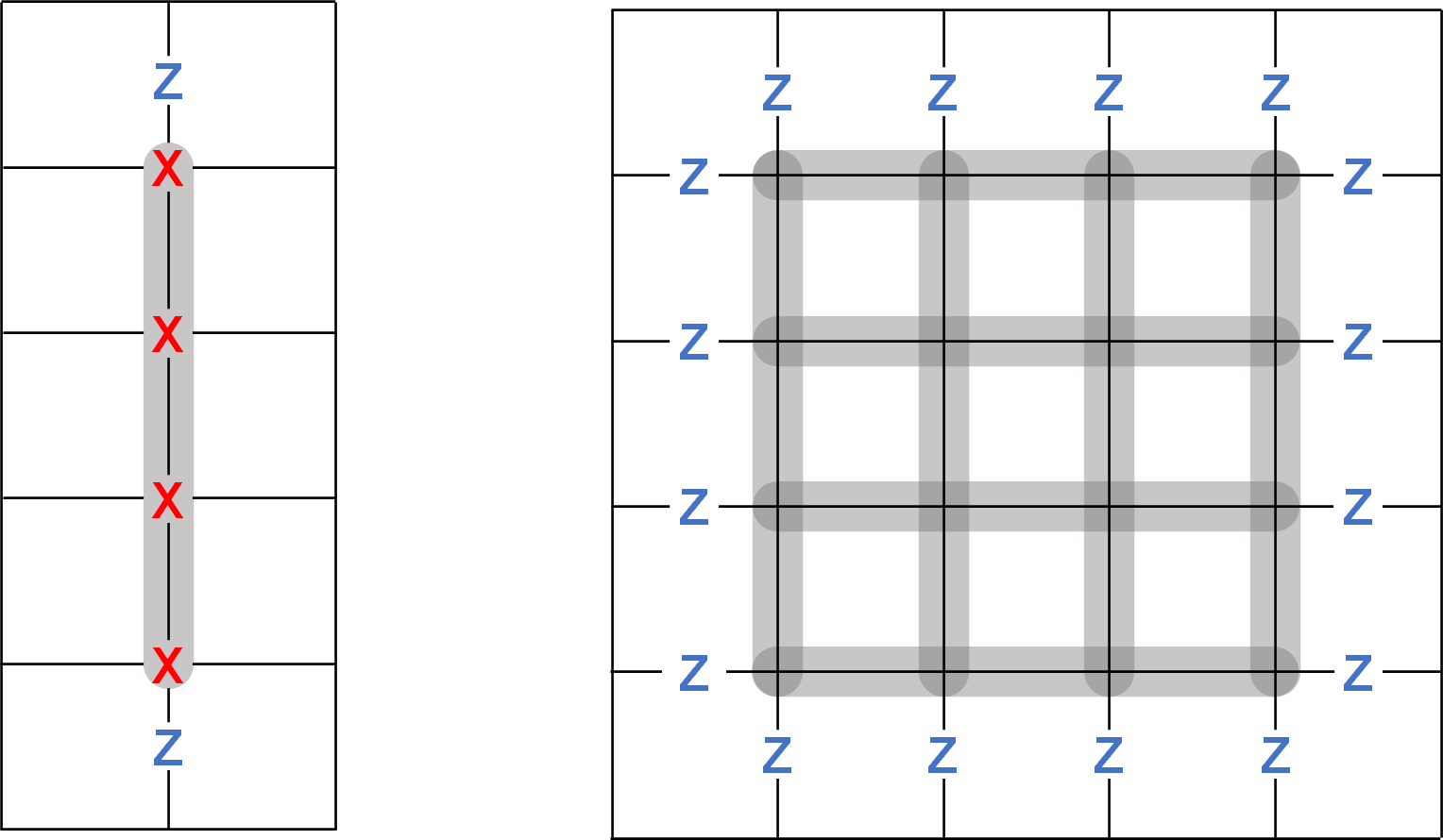}}}
\caption{(a) Hamiltonian terms of $H_\ell$ and the corresponding string operators. Stars indicate the Hamiltonian terms that are violated by a string operator. (b) Left: A truncated line symmetry $V(y_i)$. $V(x_j)$ is similar but rotated $90^\circ$. Right: The relation $V(r_\mathrm{all})$ involving all truncated lines in a $4\times 4$ region $R$. The sites within $R$ are each acted on by two line symmetries, so the bulk action cancels out and only the boundary action remains.}
\label{fig:hl}
\end{figure}

Now, we define a model which has the same $\mathbb{Z}_2$ topological order as $H_{TC}$ but exhibits fractionalization of the line symmetries. The model is defined by the following Hamiltonian,
\begin{equation}
    H_\ell = -\sum_v A^{\ell,x}_v -\sum_v A^{\ell,y}_v -\sum_p B^\ell_p.
\end{equation}
where,
\begin{equation}
\begin{aligned}
    A^{\ell,x}_v &= X_v Z_{v+\frac{\hat{x}}{2}} Z_{v-\frac{\hat{x}}{2}}, \\  A^{\ell,y}_v &= X_v Z_{v+\frac{\hat{y}}{2}}Z_{v-\frac{\hat{y}}{2}},
\end{aligned}
\end{equation}
and,
\begin{equation} \label{eq:bp}
    B^{\ell}_p = \left[\prod_{e\in p} X_e \right]\left[\prod_{v\in p} Z_e\right].
\end{equation}
The various terms in $H_\ell$ are shown in Fig.~\ref{fig:hl_terms}. Notice we have $A^{\ell,x}_vA^{\ell,y}_v=A_v$. We also observe that $H_\ell$ commutes with the subsystem symmetries $U(g)$.

$H_\ell$ can be mapped to $H_{TC}$ by a quantum circuit $\mathcal{U}_\ell$ defined as
\begin{equation}
\mathcal{U}_\ell = \prod_v CZ_{v,v+\frac{\hat{x}}{2}}CZ_{v,v-\frac{\hat{x}}{2}},
\end{equation}
where $CZ_{a,b} = |0\rangle\langle 0|_a\otimes \mathbb{1}_b + |1\rangle \langle 1 |_a \otimes Z_b$ is the two-qubit controlled-$Z$ operator. If we use the facts that $CZ_{ab} X_a CZ_{ab}^\dagger = X_aZ_b$, $CZ_{ab} Z_a CZ_{ab}^\dagger = Z_a$, and that $CZ$ is symmetric about its two inputs, we find,
\begin{equation}
    \mathcal{U}_\ell H_\ell\mathcal{U}_\ell^\dagger = -\sum_v X_v - \sum_v X_v A_v -\sum_p B_p.
\end{equation}
This Hamiltonian is clearly equivalent to $H_{TC}$ up to a local redefinition of Hamiltonian terms that preserves the ground space. The fact that $H_\ell$ and $H_{TC}$ can be related by a quantum circuit shows that they possess the same $\mathbb{Z}_2$ topological order. However, $\mathcal{U}_\ell$ is not a symmetric circuit, \textit{i.e.} it is not composed of local operations that commute with the subsystem symmetries $U(g)$. It is shown below at the end of this section that no such symmetric circuit can connect $H_{TC}$ and $H_\ell$, meaning that they are in different subsystem symmetry-enriched phases of matter.

The excitations of $H_\ell$ come in three elementary types which we call $\mathrm{e}^x$, $\mathrm{e}^y$, and $\mathrm{m}$ corresponding to violations of $A^x_v$, $A^y_v$, and $B^\ell_p$, respectively. Notice that the local operator $Z_v$ transforms $\mathrm{e}^x$ into $\mathrm{e}^y$, and vice-versa, so $\mathrm{e}^x$ and $\mathrm{e}^y$ are in the same topological sector, namely $\mathrm{e}$. However, $Z_v$ does not commute with the subsystem symmetries, and we show below that it is impossible to convert $\mathrm{e}^x$ to $\mathrm{e}^y$ with operations that respect the subsystem symmetries. 

We can construct string operators for these excitations by taking the usual toric code string operators and conjugating them by $\mathcal{U}_\ell$. The string operator $S^\mathrm{m}_\lambda$ that creates $\mathrm{m}$-anyons at its endpoints is the same as the one defined previously for the toric code, while the string operators for $\mathrm{e}^{x/y}$-anyons have the form,
\begin{equation}
    S^{\ell,\mathrm{e}}_\gamma = \prod_{e\in\gamma} X_e \prod_{v\in\mathrm{corners}} Z_v
\end{equation}
where $\gamma$ is again a path of edges on the lattice, and $Z$ operators are placed on sites wherever $\gamma$ turns a corner. The string operators are depicted in Fig.~\ref{fig:hl_terms}. $S^{\ell,\mathrm{e}}_\gamma$ creates an $\mathrm{e}^x$ ($\mathrm{e}^y$) at each endpoint site if $\gamma$ enters that site from the left or right (above or below). If we enforce the subsystem symmetries then we can only make a pair of $\mathrm{e}^x$ ($\mathrm{e}^y$) anyons along a single row (column), since the $Z_v$ operators at the corners of $S^{\ell,\mathrm{e}}_\gamma$ anti-commute with some of the symmetry generators. That is, the $\mathrm{e}^x$ ($\mathrm{e}^y$) anyons can only be moved horizontally (vertically) with symmetric operators, so we call them \textit{symmetry-protected lineons}, where we borrow the terminology lineon from the literature on fracton topological phases where it describes an anyon that can only move along a line \cite{Shirley2019frac}. These mobility restrictions violate the assumption that anyons are fully mobile that was used in Ref.~\cite{Stephen2020} and the introduction of this paper to argue against the existence of subsystem symmetry fractionalization in 2D.

\begin{figure}
    \centering
    \includegraphics[scale=0.2]{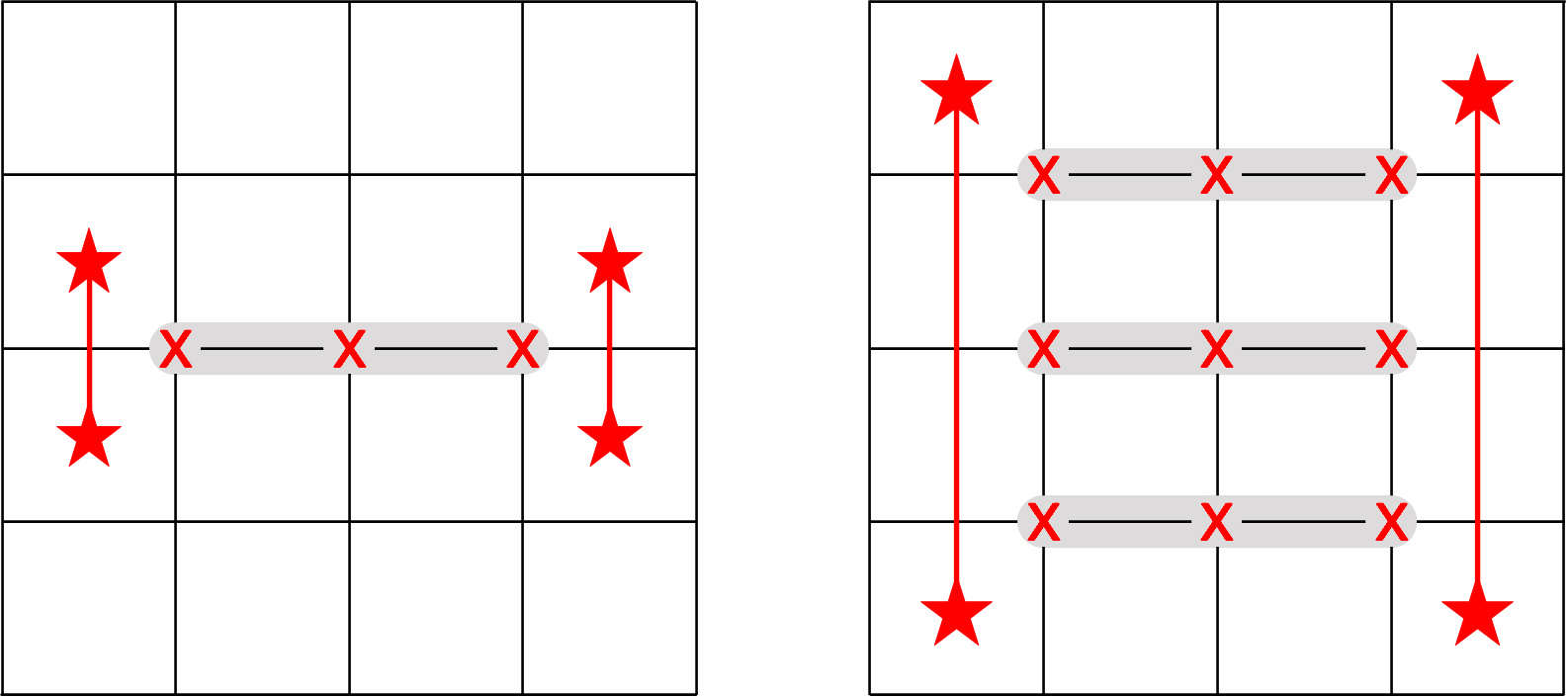}
    \caption{Truncated line symmetries create pairs of $\mathrm{m}$-anyons at the endpoints. Stacking multiple lines can separate these anyons to the corners of a rectangle.}
    \label{fig:line_defects}
\end{figure}

We now describe the nature of the symmetry fractionalization in $H_\ell$ by determining the boundary action $V(r_{\mathrm{all}})$ of the global relation $r_{\mathrm{all}}$ between all row and column symmetries.
Let $R$ be a rectangular region containing the points $v=(i,j)$ with $i\in [x_0,x_1]$ and $j\in [y_0,y_1]$. $V(x_j)$ and $V(y_i)$ can be obtained for this region as products of $A^{\ell,x}_v$ and $A^{\ell,y}_v$, respectively. Doing this, we obtain,
\begin{align}
    V(x_j) &= Z_{(x_0-\frac{1}{2},j)}Z_{(x_1 +\frac{1}{2},j)}\prod_{i=x_0}^{x_1} X_{(i,j)} \\
    V(y_i) &=Z_{(y_0 -\frac{1}{2},j)}Z_{(y_1 +\frac{1}{2},j)}\prod_{j=y_0}^{y_1} X_{(i,j)}
\end{align}
for all $i\in [x_0,x_1]$ and $j\in [y_0,y_1]$, while generators outside of $R$ simply act as the identity, see Fig.~\ref{fig:hl_rel}.  We can interpret these equations in the following way: applying the line symmetry to a finite interval creates a pair of $\mathrm{m}$-anyons at each endpoint, as shown in Fig.~\ref{fig:line_defects}, which are subsequently annihilated by the $Z$ operators at the endpoints. Taking a product of all of the localized symmetries, we obtain
\begin{multline}
    V(r_{\mathrm{all}}) \equiv \prod_j V(x_j) \prod_i V(y_i)^{-1} = \\
    \prod_{j=y_0}^{y_1} Z_{(x_0-\frac{1}{2},j)}Z_{(x_1 +\frac{1}{2},j)} \prod_{i=x_0}^{x_1} Z_{(y_0-\frac{1}{2},j)}Z_{(y_1+\frac{1}{2},j)} \text{.}
\end{multline}
This operator is equal to $S^{\mathrm{m}}_\lambda$ where $\lambda$ is a closed loop surrounding $R$, see Fig.~\ref{fig:hl_rel}. We may therefore label the fractionalization in this model by $\phi(r_{\mathrm{all}})=\mathrm{m}$. 

If the state $|\psi\rangle$ contains an $\mathrm{e}$-anyon in region $R$, we have $V(r_{\mathrm{all}})|\psi\rangle = - |\psi\rangle$ due to the mutual braiding statistics between $\mathrm{e}$ and $\mathrm{m}$ anyons. Therefore the relation between all row and column symmetries is fractionalized when acting on an $\mathrm{e}$-anyon.
This argument makes it clear that no local excitation can transform under the symmetry in this fractional manner: since $V(r_{\mathrm{all}})$ acts only on the boundary of $R$, it must act trivially on any local excitation contained in the bulk of $R$.

We can get a more fine-grained picture of the fractionalization by calculating the charge carried by a given $\mathrm{e}$-anyon under each symmetry generator. Consider a single $\mathrm{e}^x$ anyon living in row $j^*$ of the region $R$. This anyon can be created by a string operator $S^{\mathrm{e}}_\gamma$ where $\gamma$ is a horizontal line of edges on row $j^*$ with one endpoint inside $R$ and the other far outside of it. Then, we can compute the charge carried by this anyon under $U(g)$ from the commutation relation of this string operator and the localized symmetry operator $V(g)$. Doing this, we find that this anyon has a charge of $-1$ under $U(x_{j^*})$, and $+1$ under all other generators. Similarly, an $\mathrm{e}^y$-anyon in column $i^*$ has a charge of $-1$ under $U(y_{i^*})$ and $+1$ under all other generators. In both cases, the total charge under the product of all generators is $-1$, despite the fact that $\prod_j U(x_j) \prod_i U(y_i) = 1$, which again demonstrates the fractionalization.

We can show more directly that $H_\ell$ belongs to a distinct SSET phase from $H_{TC}$ by showing that they cannot be connected by a symmetric circuit. Define the operator $U_R = \prod_{v\in R} X_v$ for a rectangular region $R$. This operator can be viewed as restricting the global operator $\prod_j U(x_j)$ (or equivalently $\prod_i U(y_i)$) to the region $R$. Acting on the ground state of $H_\ell$ with $U_R$ creates $\mathrm{m}$-anyons at the corners of $R$, as shown in Fig.~\ref{fig:line_defects}. Now, Let $\mathcal{U}=\prod_i u_i$ be an arbitrary finite-depth quantum circuit, which consists of a finite number of layers of local, non-overlapping gates $u_i$ where each gate also commutes with all subsystem symmetries. Then, we observe that the local gates commute with $U_R$ everywhere except near the corners, since $U_R$ looks locally like a product of subsystem symmetries everywhere except near the corners  \footnote{For example, $U_R$ looks like a product of $U(y_i)$ near the left and right edges, and $U(x_j)$ near the top and bottom edges.}. Therefore, the operator $\mathcal{U}U_R\mathcal{U}^\dagger$ differs from $U_R$ only by operators localized near the corners of $R$. Such localized operators cannot change the topological superselection sector at each corner, so the fact that $U_R$ creates $\mathrm{m}$-anyons at its corners is unchanged by $\mathcal{U}$. Since acting with $U_R$ on the ground state of $H_{TC}$ creates (topologically) trivial excitations at its corners, we conclude that no symmetric finite-depth circuit can relate $H_\ell$ and $H_{TC}$, so they belong to distinct SSET phases.

\subsection{Relation to SSPT order} \label{sec:sspt}

We now discuss an alternative method to derive $H_\ell$ via partially gauging the symmetry of a model with nontrivial SSPT order. This is analogous to the well known method of obtaining conventional SET order by partially gauging SPT order \cite{Barkeshli2019,Garre-Rubio2017,NewSETPaper2017,Lan2019,Tantivasadakarn2021a}. In particular, the first model of SSET order in 3D was also obtained in this way \cite{Stephen2020}. 

A model of SSPT order that can be gauged to obtain $H_\ell$ is the 2D square-lattice cluster Hamiltonian $H_C$ \cite{Raussendorf2003}, which is the prototypical example of SSPT order with line-like symmetries \cite{Raussendorf2019,You2018}. It is most naturally represented on a square lattice with two qubits $A$ and $B$ per site, where line symmetries flip either all $A$ or all $B$ qubits along a given row or column. We may therefore regard the cluster Hamiltonian as having SSPT order with respect to $\mathbb{Z}_2\times\mathbb{Z}_2$ line symmetries \cite{Raussendorf2019,You2018}. 

Here, we do not need to consider the full $\mathbb{Z}_2\times\mathbb{Z}_2$ subsystem symmetry of the cluster Hamiltonian. Instead, we focus on a subgroup consisting of the $A$ line symmetries, and the $\mathbb{Z}_2$ global symmetry given by flipping all $B$ qubits (which is generated by the $B$ line symmetry). We ignore the rest of the $B$ line symmetry.
$H_\ell$ is obtained from $H_C$ by gauging this global symmetry.
To accomplish this, we add new gauge qubits on the edges of the square lattice and couple them to the original degrees of freedom via the usual minimal coupling procedure \cite{Kogut1975,Haegeman2015,Williamson2016,Shirley2019}. In this case, it is possible to use the gauge constraint to disentangle and remove the original $B$ qubits, such that we are left with the $A$ qubits on the sites of the lattice and the gauge qubits on the edges. It is straightforward to confirm that the result of this gauging procedure applied to $H_C$ is exactly $H_\ell$ where the ungauged line symmetries acting on the $A$ qubits become the fractionalized line symmetries in $H_\ell$.

The symmetry fractionalization in $H_\ell$ can be related to the SSPT order of $H_C$ before gauging, which involves an interplay of subsystem and global symmetries in 2D and has not been discussed previously, to our knowledge (although an example in 3D was discussed in Ref.~\cite{Stephen2020}). In $H_C$, acting with the $A$ symmetry in a rectangular region creates $B$ global symmetry charges at the corners of the region, and vice-versa. This fact is crucial to identifying and classifying the SSPT order \cite{You2018,Devakul2018}. After gauging, these $B$ symmetry charges become $\mathrm{m}$-anyons (in our naming convention), and we recover the important fact that acting with symmetry in a rectangular region on $H_\ell$ creates $\mathrm{m}$-anyons at the corners, which we used to prove that $H_\ell$ is in a non-trivial SSET phase.

The gauging procedure in this subsection applies equally well to the square lattice cluster model in the presence of symmetry preserving perturbations. This implies that the phase transition of the square lattice cluster model under symmetry respecting perturbations maps directly over to a phase transition of $H_\ell$ driven by the gauged perturbations. 
In a similar fashion, the model $H_\ell$ in the presence of edge perturbations was obtained in Ref.~\cite{Rayhaun2021} by stacking 1D Ising chains along rows and columns and gauging the resulting global $\mathbb{Z}_2$ symmetry that acts on all chains. This reveals that the phase transition driven by these edge perturbations is equivalent to a stack of 1D critical Ising chains coupled to a 2D $\mathbb{Z}_2$ gauge field.

\subsection{Example with $\phi(r_{\mathrm{all}})=\epsilon$}

\begin{figure}[t]
    \centering
    \subfigure[]{{\label{fig:fermion_ham}\includegraphics[scale=0.4]{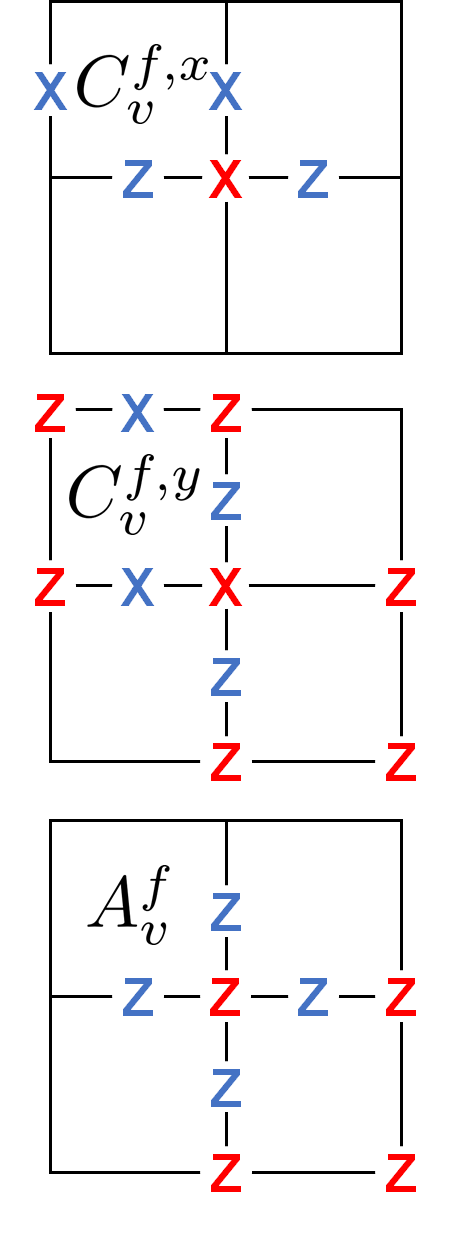}}} \hfill
    \subfigure[]{{\label{fig:fermion_ham_rel}\includegraphics[scale=0.4]{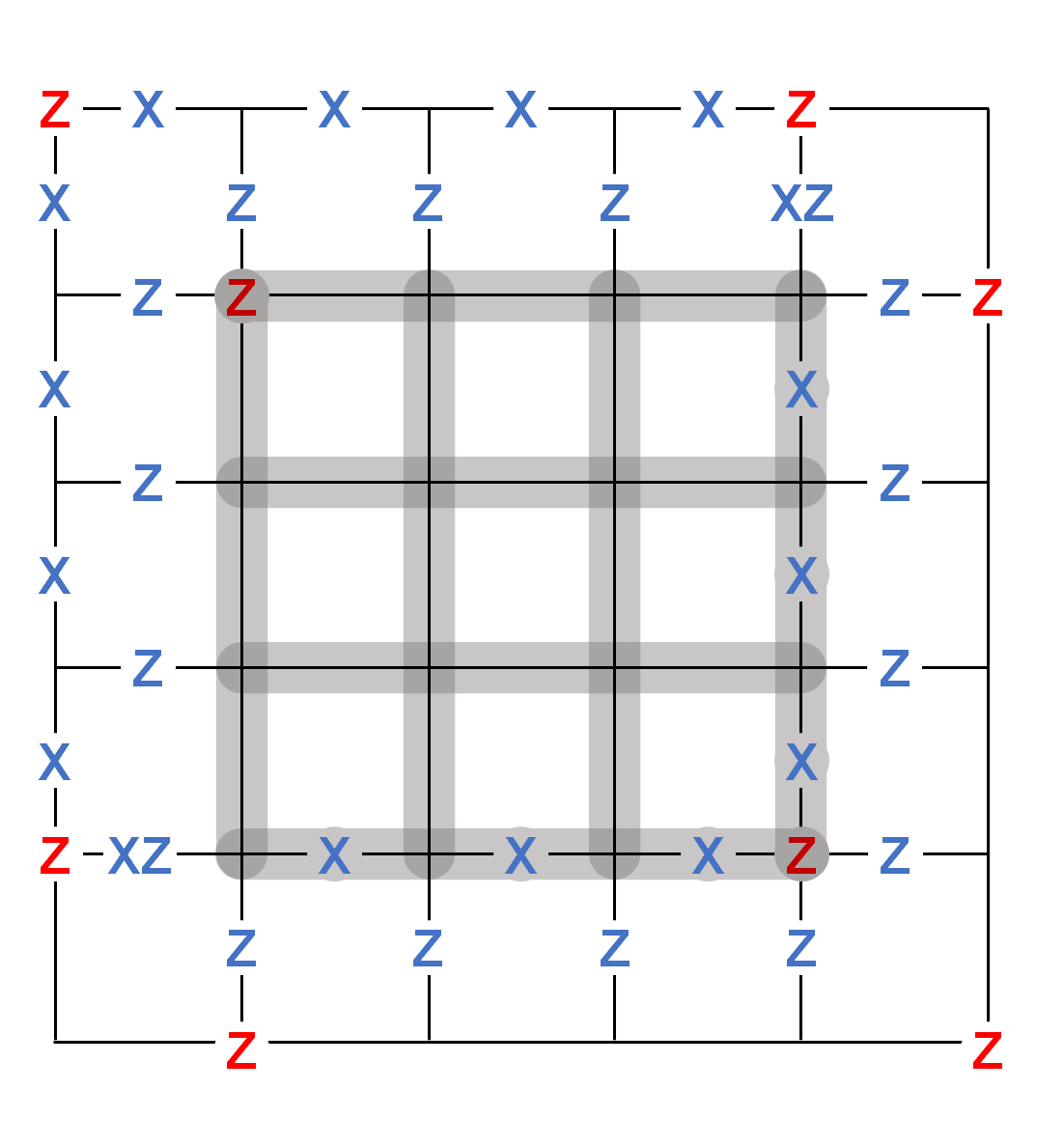}}}
    \caption{(a) The Hamiltonian terms in $H_f$. (b) The product of all line symmetries restricted to the $4\times 4$ region $R$. The boundary action is equal to the operator $V(r_{\mathrm{all}})$ which can be identified as a fermionic string operator decorated with $Z_v$ operators on the corners.}
\end{figure}

In the previous section, we gave an example where the fractionalization pattern is labelled by $\phi(r_\mathrm{all})=\mathrm{m}$. By the self-duality of the toric code topological order~\cite{Kitaev2003}, there is no meaningful distinction between this example and one where $\phi(r_\mathrm{all})=\mathrm{e}$.
In this section, we give a construction where $\phi(r_\mathrm{all})=\epsilon$. This is distinct from the other cases since $\epsilon$ is a fermion, while $\mathrm{e}$ and $\mathrm{m}$ are bosons.

We consider the following Hamiltonian,
\begin{equation}
    H_f = - \sum_v C^{f,x}_v + \sum_v C^{f,y}_v -\sum_p B^f_p
\end{equation}
where $B^f_p$ is equal to $B^\ell_p$ as defined in Eq.~(\ref{eq:bp}) and the other terms are defined in Fig.~\ref{fig:fermion_ham}. By taking a product of $C^{f,x}_v$, $C^{f,y}_v$, and $B^f_p$ where $p$ is the plaquette above and to the left of $v$, we can obtain the term $A^f_v$ pictured in Fig.~\ref{fig:fermion_ham}. $H_f$ commutes with the subsystem symmetries $U(g)$ as before. It can also be straightforwardly checked that $H_f$ can be mapped to $H_{TC}$ using a finite depth circuit consisting of $CZ$ and $CNOT$ gates where $CNOT_{a,b} = |0\rangle\langle 0|_a\otimes \mathbb{1}_b + |1\rangle \langle 1 |_a \otimes X_b$. The circuit is defined such that $C^{f,x}_v$ is mapped to $X_v$, $A^f_v$ is mapped to $A_v$ and $B^f_p$ to $B_p$. Therefore, the model has $\mathbb{Z}_2$ topological order with excitations of $A^f_v$ and $B^f_p$ corresponding to $\mathrm{e}$ and $\mathrm{m}$ anyons, respectively.

To check the fractionalization pattern, we calculate $V(r_{\mathrm{all}})$ as before. By design, a truncated line symmetry creates a pair of $\epsilon$-anyons at its endpoints, which can be seen from the fact that it anti-commutes with both $A_v^f$ and $B_p^f$ at its endpoints. The explicit localized symmetry operators $V(x_j)$ and $V(y_i)$ for the same rectangular region $R$ defined above can be obtained as products of $C^{f,x}_v$ and $C^{f,y}_v$, respectively. 
We can then calculate $V(r_{\mathrm{all}})$ which is shown in Fig.~\ref{fig:fermion_ham_rel} to be equal to a string operator for $\epsilon$ winding around the region $R$, up to additional $Z_v$ operators on the corners of $R$. This fractionalization pattern suggests that both the $\mathrm{e}$ and $\mathrm{m}$ anyons are symmetry-protected lineons, while the $\epsilon$-anyon can move freely, as can be confirmed by constructing the string operators explicitly. We note that even though the $\epsilon$ string constructed from $V(r_{\mathrm{all}})$ is decorated with $Z_v$ operators at the corners, it is possible to construct an $\epsilon$ string without such decorations, which is important for understanding that $\epsilon$-anyons are fully mobile; this can be done by taking a suitable product of $A^f_v$ and $B^f_p$ operators over a region $R$ so that all $Z_v$ operators cancel.

With this, we have demonstrated that $\phi(r_{\mathrm{all}})$ can be any anyon of the $\mathbb{Z}_2$ topological order (with $H_{TC}$ giving an example where $\phi(r_{\mathrm{all}})=1$). In Appendix \ref{app:zn}, we demonstrate the same for $\mathbb{Z}_N$ topological order. This more general setting includes cases where $\phi(r_{\mathrm{all}})$ is neither a boson nor a fermion. 
  
\section{Line symmetries in three directions} \label{sec:threelines}

The examples studied up to this point have a single global relation between the subsystem symmetry generators, and we showed how to construct models where the fractionalization class is labeled by a single anyon of the $\mathbb{Z}_2$ topological order. In this section, we study models which have line symmetries in three directions, resulting in a richer set of fractionalizable global relations, and therefore more possible patterns of fractionalization \footnote{We remark that line symmetries in three directions can also be defined with two vertex qubits per unit cell such that there is only a single global relation, we instead focus on the present case as it leads to examples with multiple global relations.}. 

To add another direction of symmetry, we consider the same horizontal and vertical symmetries $x_j$ and $y_i$ on a square lattice as before, and we now additionally consider diagonal symmetries $d_k$ along a single direction (down and to the right), where $k \in \mathbb{Z}$ labels the distinct diagonals. These symmetries are represented by,
\begin{equation}
    U(d_k) = \prod_{i} X_{(i+k,-i)} \text{.}
\end{equation}
As before, the symmetries act only on qubits residing on the square lattice sites.
The set of non-fractionalizable relations $\mathcal{R}_{nf}$ includes the first five relations of Eq.~(\ref{eq:relations}), and also
\begin{equation} \label{eq:d-relations}
\begin{aligned}
d_k^2 &= 1 \qquad \forall k\in \mathbb{Z} , \\
d_k d_l d^{-1}_k d^{-1}_l &= 1 \qquad \forall k, l\in \mathbb{Z} , \\
d_k x_j d^{-1}_k x^{-1}_j &= 1 \qquad \forall k,j \in \mathbb{Z} , \\
d_k y_i d^{-1}_k y^{-1}_i &= 1 \qquad \forall k,i \in \mathbb{Z}.
\end{aligned}
\end{equation}

The fractionalizable global relations can be labelled by a pair of bits $a,b\in \{0,1\}$ as follows,
\begin{equation} \label{eq:3relations}
    r_{a,b}\equiv \prod_j x_{2j + a} \prod_i y_{2i + b} \prod_{k} d_{2k+a+b+1} = 1 ,
\end{equation}
so that $\mathcal{R}_f = \{ r_{0,0}, r_{1,0}, r_{0,1}, r_{1,1} \}$.
For example, the relation $r_{0,0}$ says that the product of all lines on all even rows, even columns, and odd diagonals is the identity. 

As described in Sec.~\ref{sec:fractionalization}, we construct the group $\mathfrak{R}_f$ of fractionalizable relations. It should be noted that the relations $r_{a,b}$ are not all independent. That is, $\prod_{a,b = 0,1} r_{a,b} = 1$ is an identity in $\mathfrak{R}_f$. As a result, $\mathfrak{R}_f$ has three independent generators of order 2, so $\mathfrak{R}_f \cong \mathbb{Z}_2^3$.

We remark that other global relations that are more obvious at first glance are already contained in $\mathcal{R}_f$, and thus do not need to be added separately. For instance,
\begin{equation}
    r_{xy} \equiv \prod_j x_j \prod_i y_i = r_{0,0} r_{1,1} \text{.} \label{eqn:old-reln}
\end{equation} The same conclusion holds true for the product of all line symmetries in any two of the three directions; that is,
\begin{eqnarray}
r_{xd} &\equiv& \prod_j x_j \prod_k d_k = r_{0,0} r_{1,0} \\
r_{yd} &\equiv& \prod_i y_i \prod_k d_k = r_{1,0} r_{1,1} \text{.}
\end{eqnarray}
This observation gives some constraints on the possible fractionalization patterns. For example, if we wanted to construct a model where the relation $r_{xy}$ is fractionalized, as in Section~\ref{sec:line}, we must necessarily break translational invariance. By Eq.~\ref{eqn:old-reln}, the anyon $\phi(r_{xy})$ can only be non-trivial if $\phi(r_{0,0}) \neq \phi(r_{1,1})$, since $a^2 = 1$ for all $a \in \mathcal{A}$. This means even rows and columns must be distinct from odd rows and columns, so the system cannot be translationally invariant with the smallest unit cell. We construct an explicit example of this type of fractionalization that breaks translational invariance at the end of this section. It is interesting to observe that a previously allowed fractionalization type is no longer possible in a translationally invariant system, even though the additional imposition of diagonal symmetries manifestly respects translation invariance.

\begin{figure}
    \centering
    \subfigure[]{{\label{fig:3line_ham1}\includegraphics[scale=0.175]{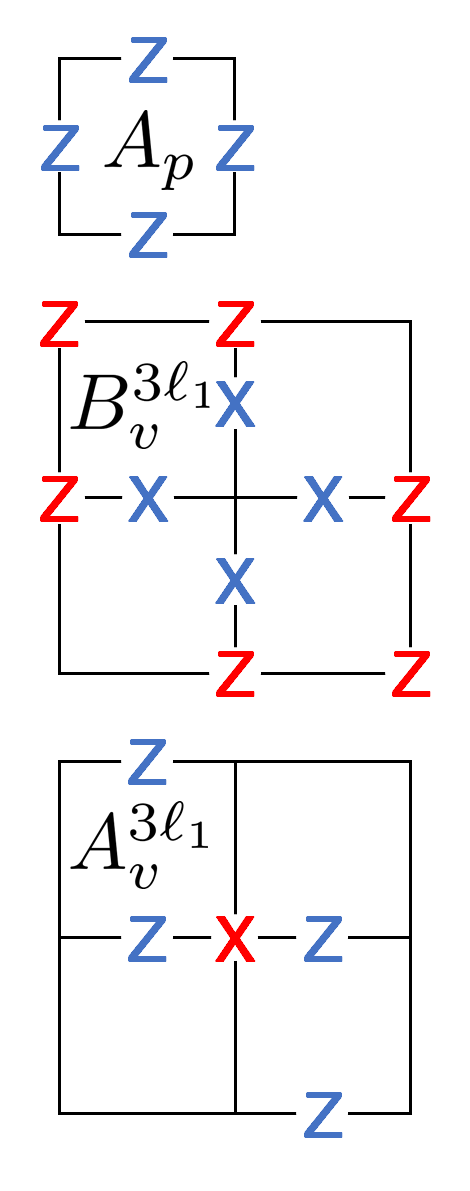}}}        \subfigure[]{{\label{fig:3line_ham1_r1}\includegraphics[scale=0.35]{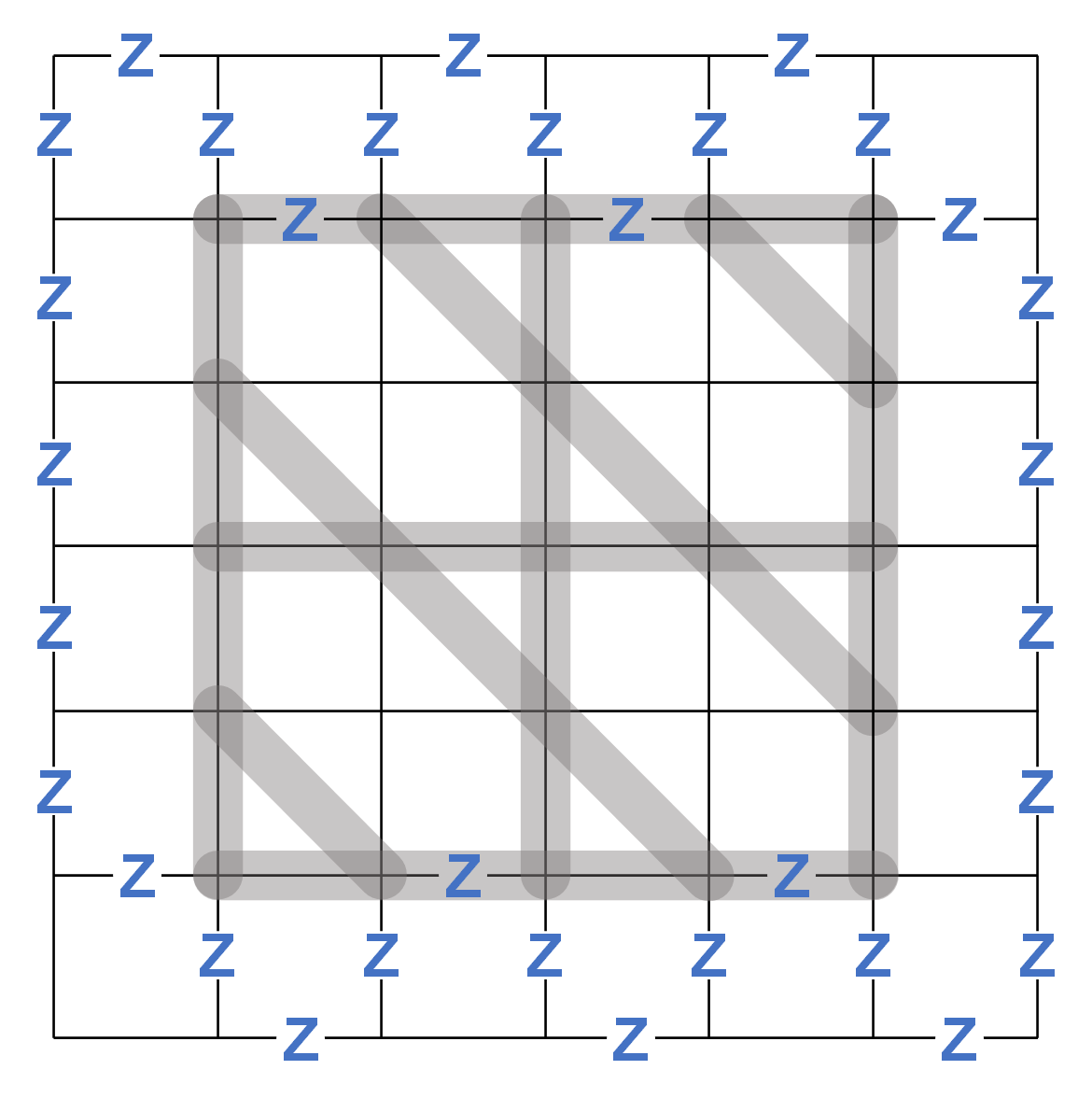}}} \hfill
    \subfigure[]{{\label{fig:3line_ham1_r2}\includegraphics[scale=0.35]{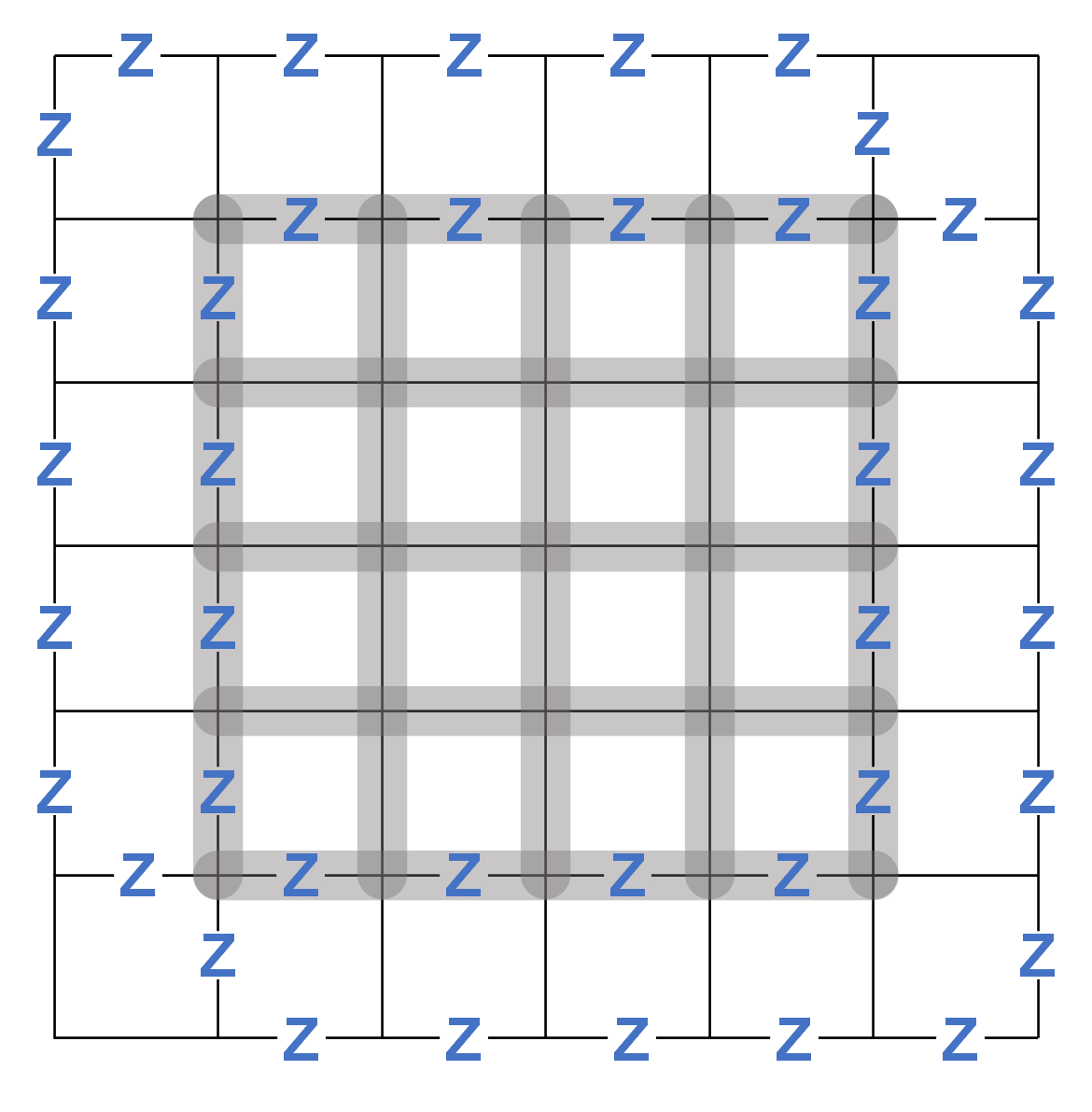}}}
    \caption{(a) The Hamiltonian terms in $H_{3\ell_1}$. (b) $V(r_{a,b})$ has the form of an $\mathrm{m}$-loop for any $a,b$, so $\phi(r_{a,b})=\mathrm{m}$. (c) $V(r_{0,0})V(r_{1,1})$ has the form of a $\mathrm{m}^2$-loop, indicating trivial fractionalization of this relation.}
\end{figure}

\subsection{Example 1: $\phi(r_{a,b})=\mathrm{m}$ $\forall a,b$.}

We begin with an example where each of the elementary relations $r_{a,b}$ has fractionalization labelled by the $\mathrm{m}$-anyon, \emph{i.e.} $\phi(r_{a,b}) = \mathrm{m}$. As a consequence of the above discussion, $r_{xy}$, $r_{xd}$ and $r_{yd}$ have trivial fractionalization since $\mathrm{m}\times\mathrm{m} = 1$.

We consider the following Hamiltonian for qubits on the edges and sites of the square lattice,
\begin{equation} \label{eq:3line_ham1}
    H_{3\ell_1} = -\sum_p A_p -\sum_v A^{3\ell_1}_v - \sum_v B^{3\ell_1}_v ,
\end{equation}
where,
\begin{equation} \label{eq:ap}
    A_p = \prod_{e\in p} Z_e,
\end{equation}
and the other terms are pictured in Fig.~\ref{fig:3line_ham1}. Note that $H_{3\ell_1}$ is naturally defined on the dual lattice compared to $H_\ell$, so $\mathrm{e}$-anyons are now associated with plaquettes rather than sites, and vice-versa for $\mathrm{m}$-anyons. The same is true for the rest of the models shown from this point on.

It is straightforward to obtain $V(g)$ for all $g=x_j,y_i,d_k$ by taking appropriate products of $A_p$ and $A^{3\ell_1}_v$. Using these, we can obtain $V(r_{a,b})$ which is the same for all $a,b$ as our model is translationally invariant. This is shown in Fig.~\ref{fig:3line_ham1_r1} where $R$ is a $5\times 5$ square. We see that $V(r_{a,b})$ is a closed string of $Z$ operators corresponding to an $\mathrm{m}$-loop, so $\phi(r_{a,b})=\mathrm{m}$ for all $a,b$. On the other hand, relations between all lines in any two directions give a trivial loop, as demonstrated in Fig.~\ref{fig:3line_ham1_r2}.

Adding a third direction of symmetry also affects the mobility of fractionalized anyons. In the case of $H_{3\ell_1}$, $\mathrm{e}$-anyons come in three types $\mathrm{e}^x$ and $\mathrm{e}^y$ and $\mathrm{e}^d$ which can move freely along only horizontal, only vertical, or only diagonal lines. These three anyon types are distinct when the subsystem symmetries are enforced.

\subsection{Example 2: $\phi(r_{0,0})=\phi(r_{1,1})=\mathrm{m}$, $\phi(r_{0,1})=\phi(r_{1,0})=1$.}

\begin{figure}
    \centering
    \includegraphics[width=\linewidth]{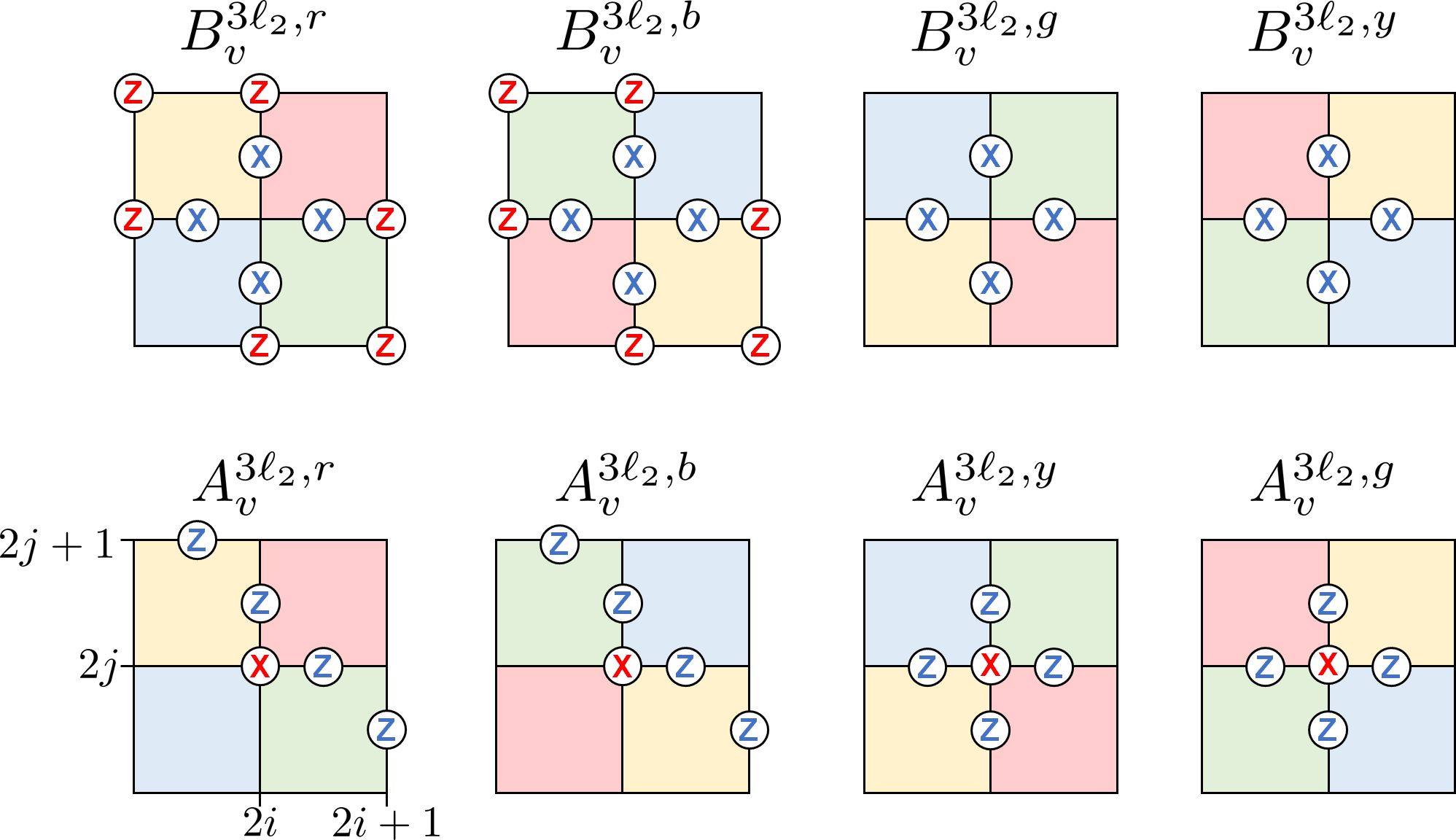}
    \caption{Definition of the colour-dependent interactions in $H_{3\ell_2}$. The site indices on the bottom left term illustrate the relation between colours and site indices, with, \textit{e.g.}, red vertices appearing on even rows and even columns.}
    \label{fig:3line_ham2}
\end{figure}
    
In this section, we construct an example where the relations $r_{xd}$ and $r_{yd}$ have non-trivial fractionalization. As discussed at the beginning of this section, this requires us to break translational invariance. We therefore label the plaquettes of the square lattice by four colours (red, blue, green, yellow) and colour the sites according to the colour of the plaquette above and to the right. Our convention relating site indices to colours is shown in Fig.~\ref{fig:3line_ham2}. We consider the following Hamiltonian,
\begin{equation} \label{eq:3line_ham2}
    H_{3\ell_2} =  - \sum_{p} A_p - \sum_{\substack{c=r,b\\ g,y}} \left( \sum_{v\in \mathcal{V}_c} B^{3\ell_2,c}_v + \sum_{v\in \mathcal{V}_c} A^{3\ell_2,c}_v  \right)
\end{equation}
where $\mathcal{V}_c$ is the set of all sites of colour $c$, $A_p$ is as defined in Eq.~(\ref{eq:ap}), and the colour-dependent terms are defined in Fig.~\ref{fig:3line_ham2}. In the same way as the previous section, it is straightforward to show that $\phi(r_{0,0})=\phi(r_{1,1})=\mathrm{m}$ while $\phi(r_{0,1})=\phi(r_{1,0})=1$. As a consequence, 
$\phi(r_{xd}) = \phi(r_{0,0}) \phi( r_{1,0} ) = \mathrm{m}$, and similarly $\phi(r_{yd}) = \mathrm{m}$. On the other hand, $\phi(r_{xy}) = 1$.

We can realize other fractionalization types in a similar manner. Observe that there are two kinds of $B_v$ terms in $H_{3\ell_2}$, one that is decorated by $Z_v$ operators and one that is not. The other fractionalization types involving only $\mathrm{m}$-anyons can be realized by assigning the decorated term to an even number of colours, and the un-decorated to the rest, and then defining $A^{3\ell_2,c}_v$ as $X_v$ along with an appropriate product of $Z$ operators on edges to ensure all terms commute. $H_{3\ell_2}$ is the result of this procedure when the red and blue sites were given the decorated term, whereas $H_{3\ell_1}$ is the result where all sites were given the decorated term.

\section{Fractal subsystem symmetries} \label{sec:fractal}

\begin{figure}[t]
\centering
\subfigure[]{{\label{fig:hf_terms}\includegraphics[scale=0.165]{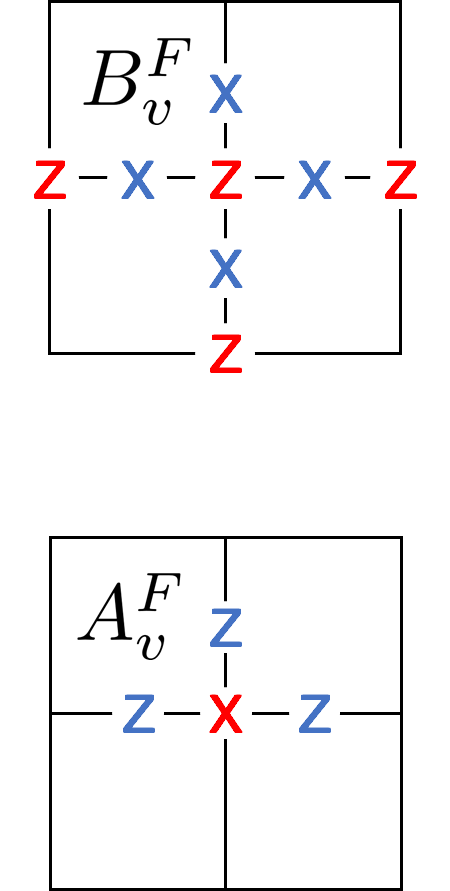}}} \hfill
\subfigure[]{\label{fig:frac_defects}\includegraphics[scale=0.33]{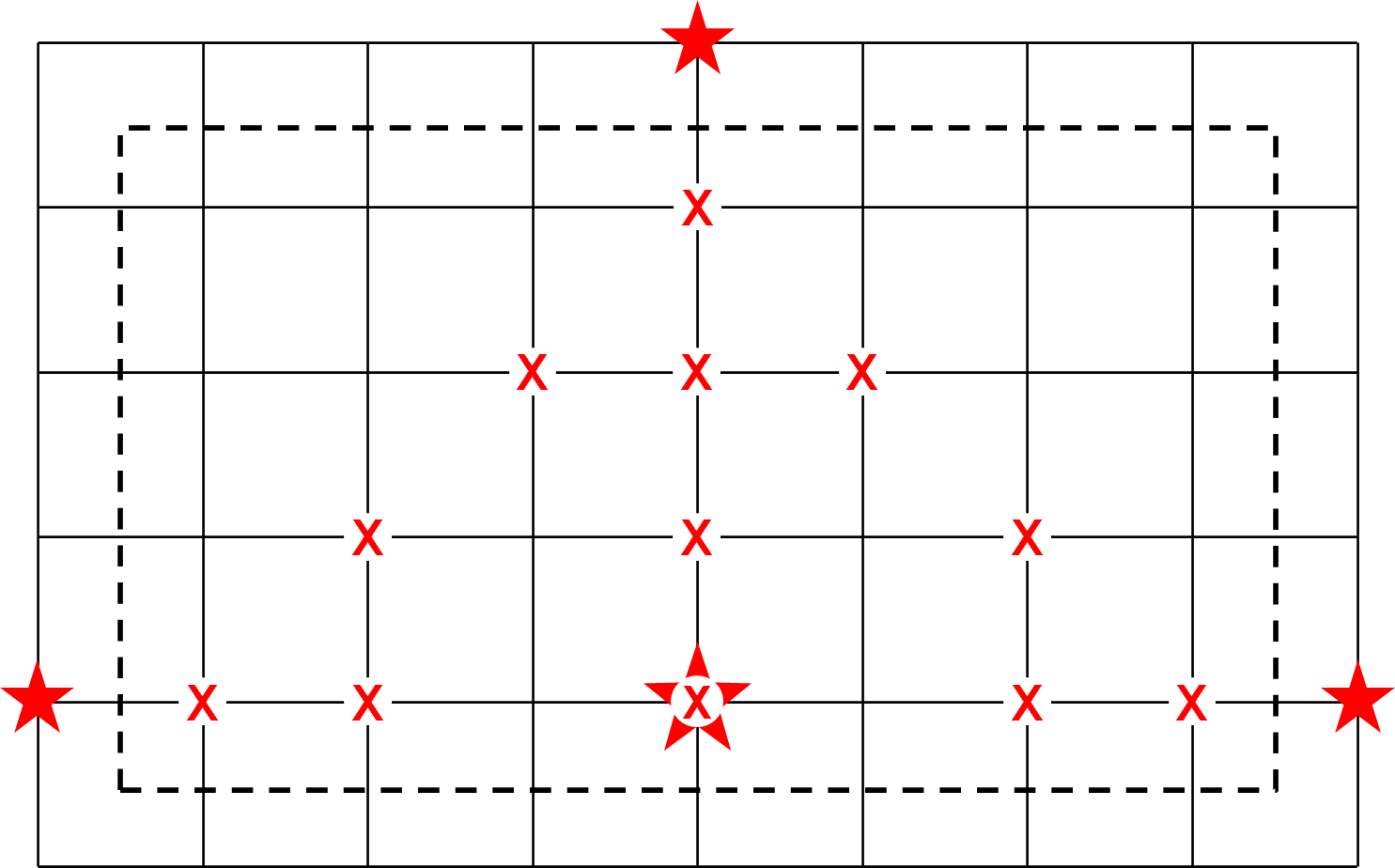}}
\subfigure[]{\label{fig:frac_logical}\includegraphics[scale=0.33]{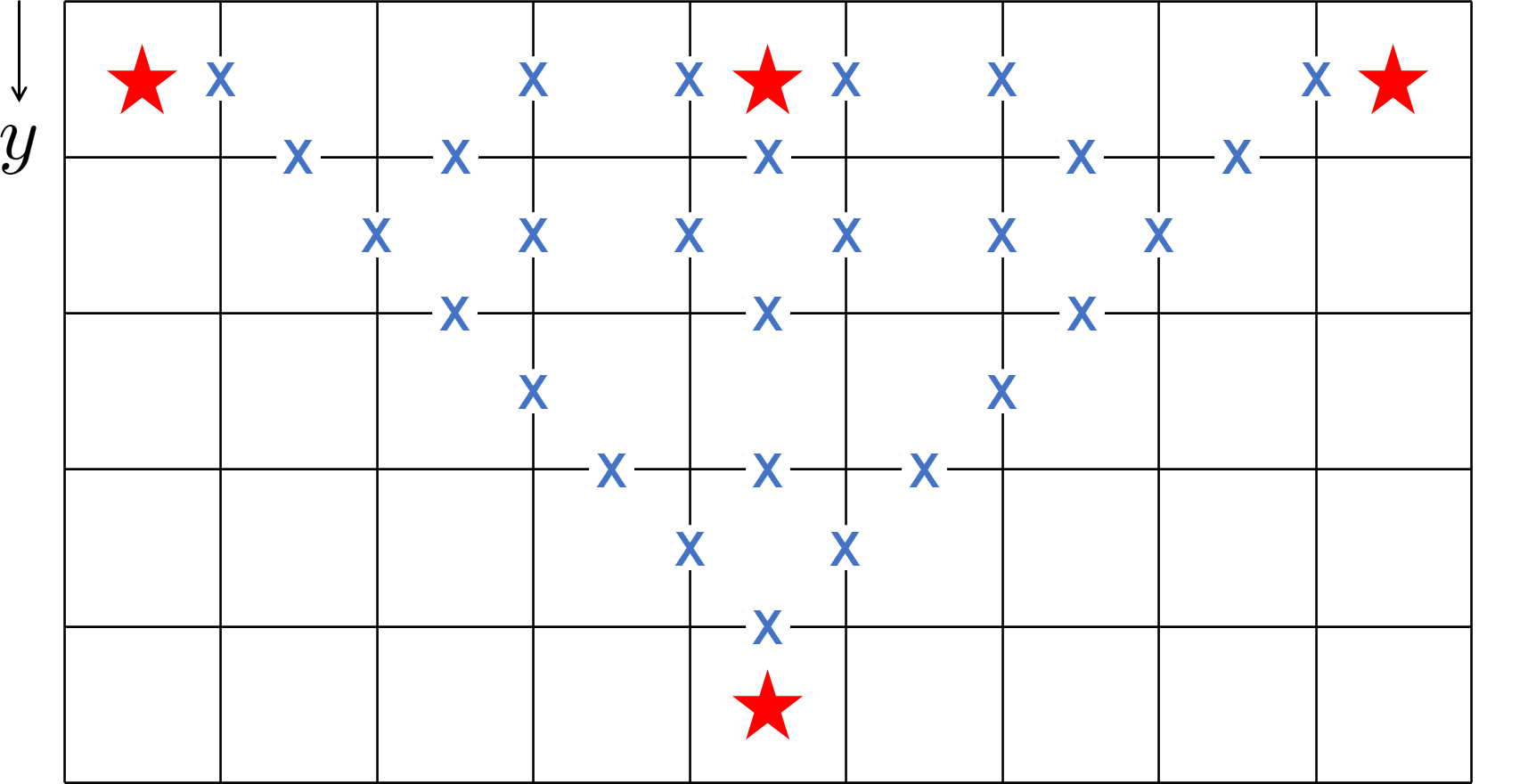}}
\caption{In these figures, the $y$-coordinate increases when moving downwards. (a) Hamiltonian terms of $H_F$. (b) A fractal symmetry truncated to the sites within the dotted rectangle. This truncation creates $\mathrm{m}$-anyons along the top and bottom of the rectangle. (c) A symmetric operator that creates $\mathrm{e}$-anyons in a fractal configuration. It can be seen that every row of horizontal edges contains an odd number of $X$ operators.}
\label{fig:hf}
\end{figure}

The above sections have dealt with subsystem symmetries generated by operators with the geometry of lines. In this section, we consider models with subsystem symmetries generated by operators that have fractal geometry. We focus on a single example for concreteness, discussing generalizations in Section \ref{sec:gfss}. Our model with fractal symmetries is defined by the following Hamiltonian for qubits on the sites and edges of the square lattice,
\begin{equation}
    H_F = -\sum_p A_p -\sum_v A^F_v  -\sum_v B^F_v
\end{equation}
where $A_p$ is as defined in Eq.~(\ref{eq:ap}),
\begin{equation}
    A^F_v = X_v Z_{v-\frac{\hat{x}}{2}} Z_{v+\frac{\hat{x}}{2}} Z_{v-\frac{\hat{y}}{2}},
\end{equation}
and,
\begin{equation}
    B^F_v = \left[ \prod_{e\ni v} X_e\right] Z_v Z_{v-\hat{x}} Z_{v+\hat{x}} Z_{v+\hat{y}}.
\end{equation}
These terms are pictured in Fig.~\ref{fig:hf_terms}. The following circuit maps $H_F$ to $H_{TC}$,
\begin{equation}
\mathcal{U}_F = \prod_v CZ_{v,v+\frac{\hat{x}}{2}}CZ_{v,v-\frac{\hat{x}}{2}}CZ_{v,v-\frac{\hat{y}}{2}},
\end{equation}
hence $H_F$ is a model of $\mathbb{Z}_2$ topological order.

To describe the subsystem symmetries of this model, we introduce the language of cellular automata (CA). We define the set binary vectors $\mbf{q}=\bigoplus_v q_v$ which have entries $q_v=0,1$ for each lattice site $v$. Then, we can define symmetry operators as,
\begin{equation}
    U(\mbf{q}) = \prod_v (X_v)^{q_v}.
\end{equation}
$U(\mbf{q})$ is a symmetry of $H_{F}$ if $\mbf{q}$ represents a valid space-time history of a certain CA. More precisely, if we define the row vectors $\mbf{q}_j=\bigoplus_i q_{(i,j)}$ and a CA matrix $f$ such that,
\begin{equation}
    (f\mbf{q}_j)_i = q_{(i,j)} + q_{(i-1,j)} + q_{(i+1,j)} \mod 2,
\end{equation}
then $U(\mbf{q})$ is a symmetry if $\mbf{q}_{j+1} = f\mbf{q}_j$ for all $j$. An example of such a symmetry operator is shown in Fig.~\ref{fig:frac_defects}. We can define a generating set of the fractal symmetries in terms of the operators $U(f_k) = U(\mbf{q}^k)$ where $\mbf{q}^k$ is a valid space-time history of the CA such that $\mbf{q}^k_{(i,0)} = \delta_{ik}$, \textit{i.e.} $U(f_k)$ contains a single $X$ in row $j=0$. Since $f$ is reversible, fixing a single row of $\mbf{q}^k$ in this way fixes all of it \footnote{There are some subtleties on how to define $\mbf{q}^{k}_j$ for $j<0$ that depend on the boundary conditions, but a valid choice can always be made \cite{Devakul2019}.}. 

While the $\mathrm{m}$-anyons (excitations of $B^F_v$) can be created in pairs and freely moved using flexible string operators composed of $Z_e$ operators, this is not the case for $\mathrm{e}$-anyons (excitations of $A_p$). In fact, no symmetric string operators exist for the $\mathrm{e}$-anyons, as opposed to the situation that occurs in $H_\ell$, where symmetric string operators without corners can be constructed. Instead, symmetric configurations of $\mathrm{e}$-anyons in $H_F$ are restricted to appear at the corners of fractal operators, as shown in Fig.~\ref{fig:frac_logical}. This implies that an isolated $\mathrm{e}$-anyon cannot move at all without violating the subsystem symmetry or creating additional excitations, analogous to fracton models such as Haah's cubic code \cite{Haah2011}, hence we call it a \textit{symmetry-protected fracton}.

More precisely, we can define the following logical operators,
\begin{equation} 
    S^{\mathrm{e}}_{\mbf{q}} = \prod_v (X_{v+\frac{\hat{x}}{2}} X_{v-\frac{\hat{y}}{2}} X_{v+\hat{x}-\frac{\hat{y}}{2}})^{q_v}.
\end{equation}
It is straightforward to check that $S^{\mathrm{e}}_{\mbf{q}}$ commutes with every term in $H_F$ whenever $\mbf{q}_{j-1} = f\mbf{q}_j$ for all $j$. It also commutes with all subsystem symmetries. This operator has a fractal geometry that is similar to that of the subsystem symmetries, except the fractal evolves ``backwards'' in space. Now, if we restrict  $S^{\mathrm{e}}_{\mbf{q}}$ to a finite strip of $y$-coordinates,
\begin{equation} \label{eq:frac_logical}
    S^{\mathrm{e}}_{\mbf{q}}(j_0,j_1) =  \prod_{\substack{v=(i,j)\\ j_0\leq j\leq j_1}} (X_{v+\frac{\hat{x}}{2}} X_{v-\frac{\hat{y}}{2}} X_{v+\hat{x}-\frac{\hat{y}}{2}})^{q_v}
\end{equation}
the resulting operator creates some symmetry-respecting configuration of $\mathrm{e}$-anyons at the top and bottom of the strip, depending on $\mbf{q}$, see Fig.~\ref{fig:frac_logical}.

\subsection{Subsystem symmetry fractionalization in $H_F$}

To understand the symmetry fractionalization in this model, we need to carefully define our symmetry group. In the context of symmetry fractionalization, we demand that the symmetries we consider are not spontaneously broken. As discussed at the end of Section~\ref{sec:fractionalization}, we can define symmetry breaking in terms of symmetry localization. We say that a symmetry $s$ is preserved if, after truncating the symmetry to a finite region $R$, the resulting excitations on $\partial R$ can be annihilated using operators supported on $E_R(s)$, defined in Section~\ref{sec:fractionalization} as the subset of $\partial R$ where the truncated and non-truncated symmetries differ locally. More colloquially, for a symmetry to be preserved, we require that excitations created by a truncated symmetry operator -- which must lie within $E_R(s)$ -- can be ``cleaned up locally.''

According to this definition, the generators $U(f_k)$ are spontaneously broken, as we now show. Truncating $U(f_k)$ to a certain finite rectangular region $R$ creates four $\mathrm{m}$-anyons on the top and bottom edges of $\partial R$ as shown in Fig.~\ref{fig:frac_defects}. In this case, $E_R(f_k)$ is contained within the top and bottom edges of $\partial R$, and there is no way to annihilate the single anyon at the top edge by acting in these areas alone, so this symmetry is spontaneously broken. In fact, for certain boundary conditions, acting with a fractal generator that spans the entire lattice can act non-trivially on the ground state subspace. In comparison, a stack of horizontal line symmetries of $H_\ell$ truncated to a rectangular region $R$ also creates anyons at its corners as seen in Fig.~\ref{fig:line_defects}. However in that case, the region $E_R(s)$ includes a pair of vertical edges, and a pair of string operators along these edges can remove the anyons, so the symmetry is preserved. This example illustrates why the question of spontaneous symmetry breaking for subsystem symmetries is more subtle than for conventional global symmetries, due to the important role played by the geometry of generators.

To construct an SSET with unbroken fractal symmetries we instead take our subsystem symmetry group to be generated by fractal \textit{pairs} represented by the group elements $p_k\equiv f_kf_{k+1}$. The corresponding operators $U(p_k)$ make pairs of $\mathrm{m}$-anyons when truncated to a finite region which can be pairwise annihilated by acting within $E_R(p_k)$, so the symmetry is preserved. The new generators $p_k$ are subject to the following relations:
\begin{equation} \label{eq:fractalrelations}
\begin{aligned}
    p_k^2 &= 1 \qquad \forall k \in \mathbb{Z} \\
    p_kp_{k'}p_k^{-1}p_{k'}^{-1} &=  1 \qquad \forall k,k'\in\mathbb{Z} \\
    \prod_k p_k &= 1
\end{aligned}
\end{equation}
Using similar reasoning as in Sec.~\ref{sec:fractionalization}, we expect that the first two relations cannot be fractionalized. In short, we can always choose the region $R$ around an anyon such that the action of $V(p_k)$ on the boundary of $R$ is sparse, meaning that we can route the string operator out of $R$ while avoiding the boundary action. On the other hand, the third relation is a fractionalizable global relation, which we again denote as $r_{\mathrm{all}}$ as it is a product of all fractal pairs, and it has non-trivial fractionalization $\phi(r_{\mathrm{all}}) = \mathrm{m}$,
as we now demonstrate.

We demonstrate the fractionalization by showing that the symmetry charge of the $\mathrm{e}$-anyon is fractional. To compute this charge, we first localize the symmetry $U(p_k)$ to the region $R_{+y}$ corresponding to the part of the plane with positive $y$-coordinate. Define ${U}_{R_{+y}}(p_k)$ to be the truncation of $U(p_k)$ where all operators acting on sites $v=(i,j)$ with $j<0$ are removed. Acting with ${U}_{R_{+y}}(p_k)$ creates two $\mathrm{m}$-anyons located at sites $v=(k,-1)$ and $v=(k+1,-1)$. These excitations can be annihilated using the operator $Z_{(k+\frac{1}{2},-1)}$ located on the edge between the two anyons. Therefore we can define the localized symmetry operator for the upper-half plane as ${V(p_k)=Z_{(k+\frac{1}{2},-1)}{U}_{R_{+y}}(p_k)}$.

Now consider a single $\mathrm{e}$-anyon living on a site ${v=(i,j^*)}$ with $j^*>0$. We can create this anyon using a fractal operator of the form defined in Eq.~(\ref{eq:frac_logical}) and shown in Fig.~\ref{fig:frac_logical}. This operator anticommutes with $V(p_k)$ if it contains an $X$ operator at position ${(k+\frac{1}{2},-1)}$, and otherwise it commutes. Due to the fact that the cellular automaton that defines our model has an odd number of terms, there is always an odd number of $X$ operators appearing on the horizontal edges in any row of $S^{\mathrm{e}}_{\mbf{q}}$, as can be seen in Fig.~\ref{fig:frac_logical}. Therefore, this $\mathrm{e}$-anyon has a $-1$ charge under an odd number of generators $U(p_k)$. This contradicts the fact that $\prod_k U(p_k)=\mathbb{1}$, which demonstrates the fractionalization of this global relation. Since the symmetry is fractionalized acting on the $\mathrm{e}$-anyon, we  conclude that $\phi(r_{\mathrm{all}})=\mathrm{m}$. This can be verified directly by taking products of truncated fractal pairs. However, this is more cumbersome than the previous cases due to the complicated structure of the fractal symmetries.

\subsection{Generalized fractal subsystem symmetries}
\label{sec:gfss}

In Section \ref{sec:sspt}, we pointed out the fact that $H_\ell$ can be obtained by partially gauging the symmetry of a model with SSPT order. The same is true for $H_F$, and we use this observation to define an infinite family of models with different geometries of fractal symmetry. In the same way that $H_\ell$ can be obtained by gauging the square-lattice cluster Hamiltonian, $H_F$ can be obtained by gauging the Fibonacci cluster Hamiltonian defined in Refs.~\cite{Devakul2018a,Devakul2019}. This Hamiltonian again has two qubits per unit cell and has fractal subsystem symmetries which generate global symmetries that flip all $A$ or all $B$ qubits. If we gauge the global symmetry on the $B$ qubits, the resulting model is exactly $H_F$. In Ref.~\cite{Devakul2019}, an infinite family of cluster states with SSPT order under fractal symmetries were defined using cellular automata, with the model that results in $H_F$ upon gauging being one example. It is straightforward to check that the above analysis applies to all of the models in that family which have a bipartite global symmetry, which is true if and only if the defining cellular automata has an odd number of non-zero entries \footnote{Interestingly the Sierpinkski cluster Hamiltonian, \textit{i.e.} the cluster Hamiltonian on a honeycomb lattice, which is arguably the simplest cluster Hamiltonian with fractal symmetries, does not fulfill this condition.}. In such cases we can gauge the global symmetry on one sublattice to construct another model where the global relation among fractal pair generators has non-trivial fractionalization.

\begin{figure}[t]
    \centering
    \includegraphics[scale=0.2]{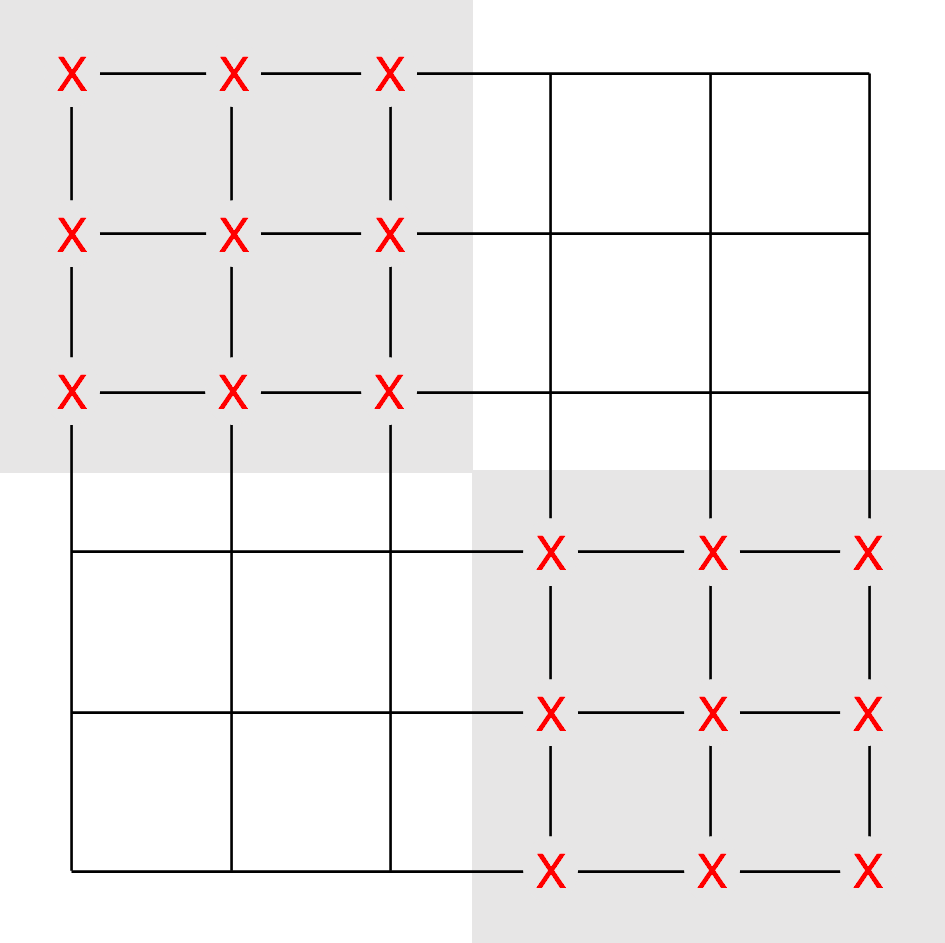}
    \caption{The cone symmetries of $H_\ell$ which can be viewed as simple fractal symmetries moving diagonally.}
    \label{fig:cone_symms}
\end{figure}

It is interesting to observe that our model $H_\ell$ with line symmetries can also be understood in this framework, despite the fact that the line symmetries seem geometrically distinct from fractal symmetries. The square lattice cluster state $H_C$ can also be defined in terms of cellular automata on a $45^\circ$-rotated square lattice, although this requires either the use of quantum cellular automata \cite{Stephen2019a} or cellular automata where each cell contains two qubits rather than one. In either case, the natural fractal subsystem symmetries that arise are cone-like symmetries pictured in Fig.~\ref{fig:cone_symms}, which one can straightforwardly verify are indeed symmetries of $H_\ell$. These cone symmetries are spontaneously broken in the same way as the fractal generators $U(f_k)$ \footnote{In the case of cone symmetries, the $\mathrm{m}$-anyons created at the corners of a rectangular region cannot be paired up as in Fig.~\ref{fig:line_defects} since the region $E_R(s)$ has changed.}. The preserved line symmetries defined in Eq.~\ref{eq:line_symms} are created by multiplying pairs of neighbouring cones, \textit{e.g.} multiplying the operator in Fig.~\ref{fig:cone_symms} with its translate one site right (down) gives a line symmetry acting on a column (row). Therefore, the fractionalization of the line symmetries in $H_\ell$ can also be understood in terms of fractionalization of fractal pairs like we saw in $H_F$, which helps to unite the phenomena in these two models. It is interesting to note that all of our examples respect a larger subsystem symmetry group that is spontaneously broken (which can be seen explicitly with an appropriate choice of boundary conditions), and we essentially restrict to a subgroup that is not broken to attain the desired behaviour.

\section{Outlook} \label{sec:outlook}

In this paper, we have studied fractionalization of subsystem symmetries in $\mathbb{Z}_2$ topological order (with some straightforward generalizations to $\mathbb{Z}_N$ given in Appendix \ref{app:zn}). We showed that the fractionalization class in each of our examples can be labelled by a map from a group of global relations among symmetry generators to the fusion group of abelian anyons. 
An evident future direction is to extend our analysis to more general types of topological order in 2D. As a first step, our results for fractionalization in $\mathbb{Z}_N$ topological order should straightforwardly extend to any finite abelian group $G$. From here, one can use the results of Ref.~\cite{Ellison2021} to obtain fractionalized models with arbitrary nonchiral abelian topological order, equivalent to the abelian twisted quantum double models. One topic that was not covered in this paper is the idea of anomalous symmetry fractionalization. In the case of global symmetries, certain symmetry fractionalization patterns are anomalous, meaning that they cannot be realized in strictly 2D, and must instead appear at the boundary of a 3D bulk system \cite{Vishwanath2013,Wang2013,Chen2015}. At the moment, it is not yet clear if there exist anomalous patterns of subsystem symmetry fractionalization in 2D.

Our findings are also  relevant for topological order in higher dimensions. While certain types of subsystem symmetry fractionalization were already known to be possible in 3D \cite{You2020,Stephen2020}, the concept of fractionalization of a global relation has yet to be explored in 3D. Some first steps in this direction can be taken by using the relationship between SSPT and SSET order. Namely, by appropriately gauging global symmetry subgroups of known SSPT orders in 3D, one can obtain models of 3D SSET order, as was done in Ref.~\cite{Stephen2020}. For example, by partially gauging the 3D model with line symmetries defined in Ref.~\cite{You2018}, one can obtain an interesting model where the allowed geometry of loop excitations is restricted in the presence of subsystem symmetries. We expect that the richness of topological order in 3D and higher, which includes fracton behaviour and higher-dimensional excitations, should lead to an equally rich landscape of possible types of subsystem symmetry fractionalization.

In this work we have focused on the fractionalization of global relations, a natural question is whether a similar phenomenon can occur for local relations. Such relations are familiar from the context of higher form symmetries, where a $k$-form symmetry satisfies a $(k-1)$-form relation. Preliminary reasoning points to such fractionalization neccesarily being trivial, as a truncated $(k-1)$-form relation applies an element of the $k$-form symmetry to the boundary introduced by truncation. If the $k$-form symmetry is respected, such an operator neccesarily corresponds to a trivial superselection sector. This differs from the examples we have introduced in this work where the boundary operators associated to truncated relations correspond to topologically nontrivial excitations.

Finally, we comment on the relationship between anyon condensation and subsystem symmetry fractionalization. In the case of global symmetries, it is known that condensing a fractionalized anyon necessarily requires breaking the corresponding symmetry (otherwise the ground state would not transform in a well-defined way under the symmetry) which has interesting effects on phase transitions out of the phase \cite{Sun2018,Wang2018,Bischoff2019,Garre-Rubio2021}. In the case of subsystem symmetries, we would also expect that symmetry breaking is required upon condensing the fractionalized anyon. Investigating such condensation transitions and their relations to previously observed forms of sub-dimensional criticality \cite{Lake2021,Rayhaun2021} remains an interesting open problem.

\vspace{.5cm}

\section*{Acknowledgements}
DTS thanks Marvin Qi for helpful discussions. The research of MH is supported by the U.S. Department of Energy, Office of Science, Basic Energy Sciences (BES) under Award number DE-SC0014415. This work was also partly supported by the Simons Collaboration on Ultra-Quantum Matter, which is a grant from the Simons Foundation (651440, MH, DS; 651438, AD), and the Simons Collaboration on It from Qubit (DJW). It was also supported by the Institute for Quantum Information and Matter, an NSF Physics Frontiers Center (PHY-1733907, AD). JGR has been partially supported by the ERC under the European Union’s Horizon 2020 research and innovation programme through the ERC-CoG SEQUAM (Grant Agreement No. 863476)

\bibliographystyle{apsrev4-1}
\bibliography{sset_biblio}

\appendix

\section{Obtaining the fractionalization class in terms of generators and relations}
\label{app:omega}

In Section~\ref{sec:fractionalization}, in reviewing symmetry fractionalization for global symmetries, we described the symmetry group $G$ in terms of generators and relations, with the information about symmetry fractionalization encoded in a homomorphism $\phi : \mathfrak{R} \to {\cal A}$, where $\mathfrak{R}$ is the group of relations defined in Section~\ref{sec:fractionalization} and ${\cal A}$ is the fusion group of abelian anyons. In this appendix, we show how to obtain a 2-cocycle $\omega$, and thus a fractionalization class $[\omega] \in H^2(G, {\cal A})$, given the homomorphism $\phi$.

We first recall some definitions from Section~\ref{sec:fractionalization}. We let $\mathcal{S}$ be a set of generators, and $\mathcal{R}$ a set of relations (\emph{i.e.} finite products of generators and their inverses). The group $G$ is given as a quotient $G = F(S) / \mathfrak{R}$, where $F(S)$ is the free group over $S$, and the group of generators $\mathfrak{R}$ is the smallest normal subgroup of $F(S)$ that contains $\mathcal{R}$.

Elements of $G$ are cosets of $\mathfrak{R}$ in $F(S)$, and for each $g \in G$ we choose a representative $\Gamma(g) \in F(S)$ such that $\Gamma(g)$ lies in the coset $g$. That is, $\Gamma(g)$ is a choice of presentation of $g$ as a product of generators. We clearly have
\begin{equation}
\Gamma(g_1) \Gamma(g_2) = r(g_1, g_2) \Gamma(g_1 g_2) \text{,}
\end{equation}
where $r(g_1, g_2) \in \mathfrak{R}$. The function $r(g_1, g_2)$ satisfies a cocycle condition obtained from associativity of multiplication by considering the product $\Gamma(g_1) \Gamma(g_2) \Gamma(g_3)$, namely
\begin{equation}
r(g_1, g_2) r(g_1 g_2, g_3) = ^{\Gamma(g_1)}r(g_2, g_3)  r(g_1, g_2 g_3) \text{.} \label{eqn:r-cocycle}
\end{equation}
Here, the superscript denotes conjugation, \emph{i.e.} $^{\Gamma(g_1)}r(g_2, g_3) = \Gamma(g_1) r(g_2, g_3) \Gamma(g_1)^{-1}$.
Moreover if we redefine $\Gamma(g) \to \Lambda(g) \Gamma(g)$ for $\Lambda(g) \in \mathfrak{R}$, we have
\begin{equation}
r(g_1, g_2) \to 
^{\Gamma(g_1)} \Lambda(g_2)^{-1} \Lambda(g_1)^{-1} r(g_1, g_2) \Lambda(g_1 g_2)   \label{eqn:r-transf} 
\end{equation}

We define $\omega(g_1, g_2) \in {\cal A}$ by
\begin{equation}
\omega(g_1, g_2) = \phi(r(g_1, g_2)) \text{.}
\end{equation}
We would like to show that $\omega$ thus defined satisfies the 2-cocycle condition, and moreover is multiplied by a 2-coboundary under the transformation given in Eq.~\ref{eqn:r-transf}.
This is easily seen to hold by applying $\phi$ to Eq.~\ref{eqn:r-cocycle} and Eq.~\ref{eqn:r-transf}, provided that the homomorphism $\phi$ satisfies the property
\begin{equation}
\phi(r) = \phi( f r f^{-1}) \text{,} \label{eqn:phiprop}
\end{equation}
for arbitrary $r \in \mathfrak{R}$ and $f \in F(S)$, which we now establish. The element $f$ is a product of generators and their inverses, so let $V_R(f)$ be the corresponding product of symmetry localizations $V_R(s)$. Now for a relation $r = s_1 \cdots s_n = 1$, we have
\begin{equation}
V_R(s_1) \cdots V_R(s_n) = V_R(r) \text{,}
\end{equation}
where $V_R(r)$ is an abelian anyon string operator on $\partial R$. Conjugating both sides by $V_R(f)$ we have
\begin{equation}
V_R(f) V_R(s_1) \cdots V_R(s_n) V_R(f)^{-1} = V_R(r) \text{,}
\end{equation}
where the right-hand side is unchanged because $V_R(f)$ commutes with abelian anyon string operators on $\partial R$. This gives the desired property Eq.~\ref{eqn:phiprop}.

\section{Subsystem symmetries do not permute anyon types}
\label{app:permute}

Here we give a simple argument that subsystem symmetries cannot realize non-trivial permutations of anyon types. The argument proceeds by contradiction. Let $U(s)$ be a subsystem symmetry, and let $| \psi_a \rangle$ be a state with an $a$-anyon located within $\operatorname{Supp} U(s)$. We can work in an infinite system with no other excitations present, or in a finite system where all other excitations are far away, with spatial locations such that they are not transformed by any of the operators introduced, and thus can be ignored. We assume that in the state $| \psi_b \rangle = U(s) | \psi_a \rangle$, the $a$-anyon has transformed into a $b$-anyon.

Let $S$ be an anyon string operator such that in that state $S_a | \psi_a \rangle$, the $a$-anyon has been moved out of $\operatorname{Supp} U(s)$. This is possible for any subsystem symmetry, including for instance both linear and fractal symmetries. Note that $S$ is supported in a bounded region containing the original and final locations of $a$.

Now we consider the state $U(s) S | \psi_a \rangle$, which contains an $a$-anyon, because the $a$-anyon in $S | \psi_a \rangle$ is not acted on by $U(s)$. However we also have
\begin{equation}
U(s) S | \psi_a \rangle = [ U(s) S U(s)^\dagger] | \psi_b \rangle \text{.}
\end{equation}
Because $[ U(s) S U(s)^\dagger]$ has bounded support, the state on the right-hand side instead contains a $b$-anyon, and we have a contradiction.

\section{Fractionalization in $\mathbb{Z}_N$ topological order} \label{app:zn}

\begin{figure}
    \centering
    \subfigure[]{{\label{fig:zn_tc}\includegraphics[scale=0.45]{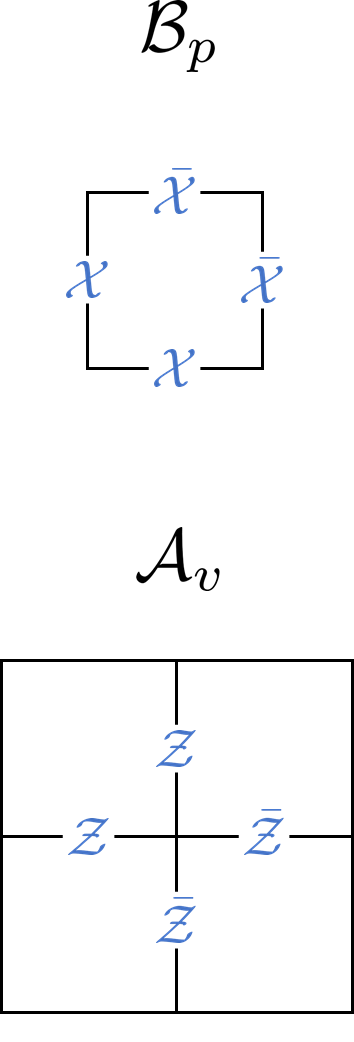}}}    \hfill \subfigure[]{{\label{fig:hab_terms}\includegraphics[scale=0.45]{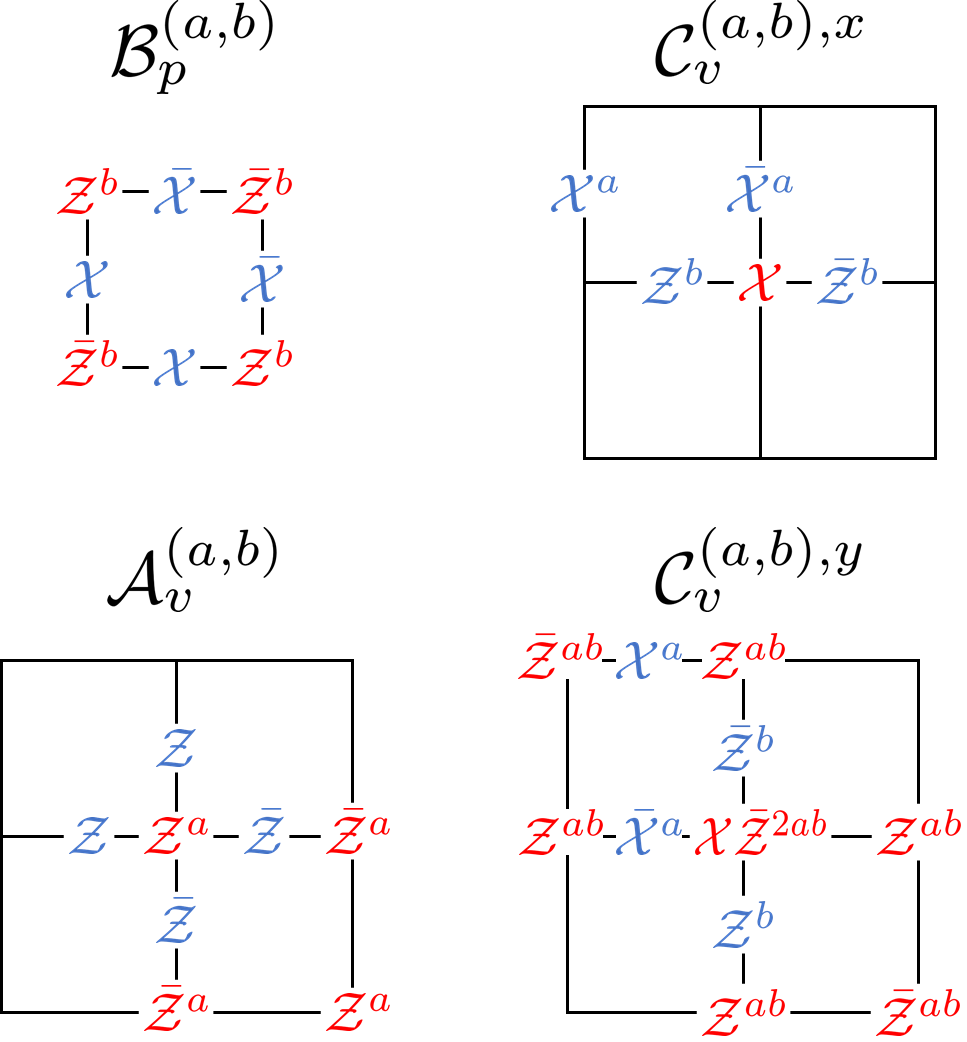}}}
    \subfigure[]{{\label{fig:hab_rel}\includegraphics[scale=0.45]{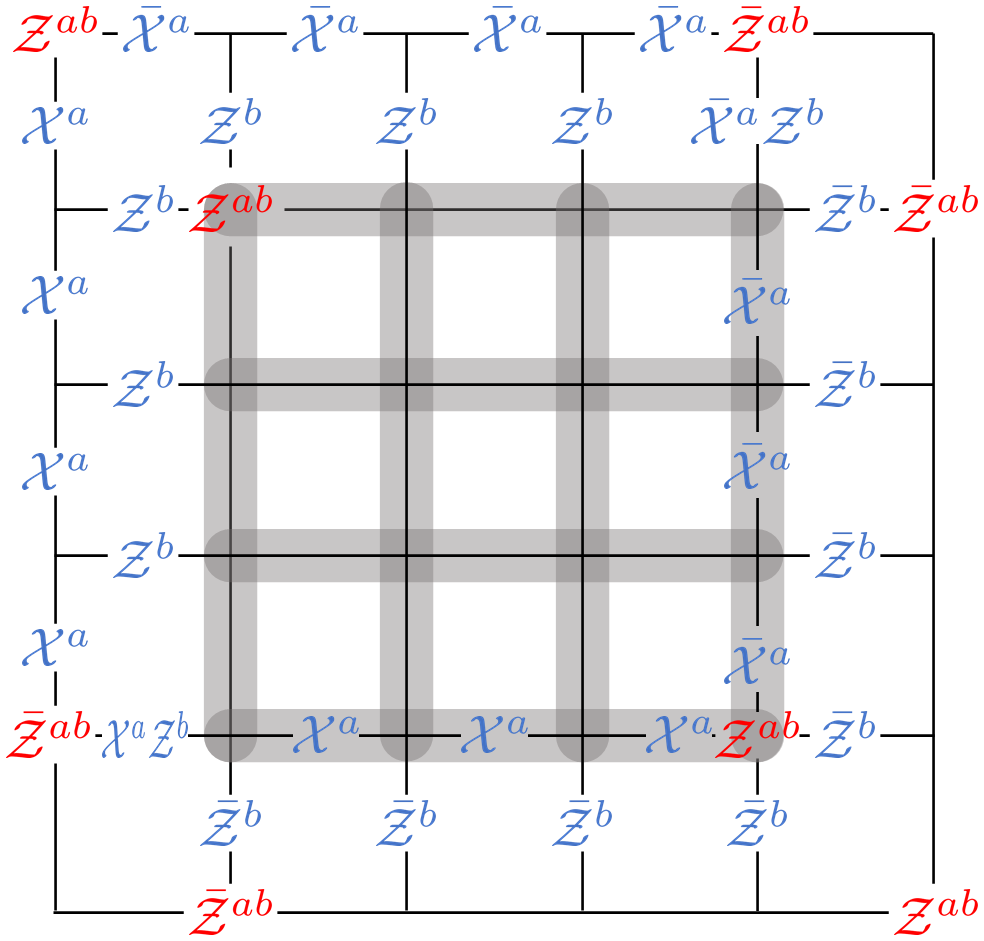}}}
    \caption{(a) The Hamiltonian terms in $H_{TC_N}$. We use the notation $\bar{\mathcal{O}} \equiv \mathcal{O}^{-1}$ for compactness. (b) The Hamiltonian terms of $H_{(a,b)}$. (c) $V(r_{\mathrm{all}})$ has the form of a loop of $\epsilon_{a,b}$, so $\phi(r_{\mathrm{all}})=\epsilon_{a,b}$.}
\end{figure}

In this section, we generalize the results in Section \ref{sec:line} to $\mathbb{Z}_N$ topological order, which is similar to $\mathbb{Z}_2$ topological order in that $\mathrm{e}$ and $\mathrm{m}$ form a generating set of anyons, except now these anyons have order $N$, $\mathrm{e}^N=\mathrm{m}^N=1$. An arbitrary anyon in $\mathrm{Z}_N$ topological order can therefore be labelled by two mod-$N$ integers $a,b$ such that ${\epsilon_{a,b} = \mathrm{e}^a\mathrm{m}^b}$. Then the braiding statistic between two arbitrary anyons $\epsilon_{a,b}$ and $\epsilon_{a',b'}$ is $\omega^{ab'+ba'}$ where $\omega=e^{2\pi i/N}$  and $\epsilon_{a,b}$ has a statistical angle $\omega^{ab}$. This statistical angle is in general different from $\pm 1$, so a general $\epsilon_{a,b}$ is neither a boson nor a fermion. 

The $\mathbb{Z}_N$ topological order is realized by the generalized $\mathbb{Z}_N$ toric code. This model is defined on a lattice with $N$-dimensional on-site Hilbert space spanned by the states $\{|0\rangle,\dots, |N-1\rangle\}$ for which we define the generalized $X$ and $Z$ operators $\mathcal{X} = \sum_{d=0}^{N-1} |d+1\rangle \langle d|$ and $\mathcal{Z}=\sum_{d=1}^{N-1} \omega^d |d\rangle\langle d|$. The Hamiltonian is then,
\begin{equation}
    H_{TC_N} = -\sum_v \mathcal{X}_v - \sum_{v} \mathcal{A}_v - \sum_p \mathcal{B}_p + h.\,c.
\end{equation}
whose terms are depicted in Fig.~\ref{fig:zn_tc}. The anyon $\mathrm{e}$ corresponds to an excitation $\mathcal{A}_v = \omega$, while $\mathrm{m}$ corresponds to $\mathcal{B}_p=\omega$. This Hamiltonian commutes with $\mathbb{Z}_N$ line symmetries in two directions. These are generated by elements $x_j$ and $y_i$ which satisfy the same relations as in Eq.~(\ref{eq:relations}), except the first two lines are replaced by $x_j^N=1$ and $y_i^N=1$. Such a subsystem symmetry can be represented by the following operators,
\begin{equation} \label{eq:line_symms_N}
    U(x_j) = \prod_{i} \mathcal{X}_{(i,j)}, \quad U(y_i) = \prod_{j} \mathcal{X}_{(i,j)}.
\end{equation}

We define a family of models having $\mathbb{Z}_N$ topological order which fractionalize the $\mathbb{Z}_N$ line symmetries according to $\phi(r_{\mathrm{all}})=\epsilon_{a,b}$ as follows,
\begin{equation}
\begin{aligned}
    H_{{(a,b)}} =
    &-\sum_v  \mathcal{A}^{(a,b)}_v -\sum_p \mathcal{B}^{(a,b)}_p \\
    &-\sum_v  \mathcal{C}^{(a,b),x}_v - \omega^{ab} \sum_v \mathcal{C}^{(a,b),y}_v + h.\, c.,
\end{aligned}
\end{equation}
where the various terms are depicted in Fig.~\ref{fig:hab_terms}. Here, we explicitly include $\mathcal{A}_v^{(a,b)}$ in the Hamiltonian as it cannot be generated from the other three terms for general values of $a$ and $b$. $H_{(a,b)}$ can be constructed with the following principle in mind. By coupling the terms $\mathcal{A}_v$ and $\mathcal{B}_p$ in $H_{TC_N}$ to symmetry charges $\mathcal{Z}^a$ and $\mathcal{Z}^b$ we create the terms $\mathcal{A}^{(a,b)}_v$ and $\mathcal{B}^{(a,b)}_p$, respectively. Because of this, a truncated line symmetry creates a pair of $\epsilon_{a,b}$ and its inverse anyon at each endpoint. $V(r_{\mathrm{all}})$ then multiplies the truncated lines such that these anyons are stitched together to create a closed loop, as shown in Fig.~\ref{fig:hab_rel}. We therefore have $\phi(r_{\mathrm{all}})=\epsilon_{a,b}$.

One can readily observe that both $H_\ell$ and $H_f$ defined in the main text are examples of $H_{(a,b)}$, where $N=2$ and $(a,b)=(0,1)$ and $(1,1)$, respectively. This construction also straightforwardly generalizes to any topological order based on a finite abelian group $G$, as captured by the $G$-quantum double model \cite{Kitaev2003}.

\end{document}